 \documentclass[final,5p,times,twocolumn]{elsarticle}

\usepackage{graphicx}
\graphicspath{{./}} 

\usepackage{tabularx}

\usepackage{amssymb}
\usepackage{amsthm}

\begin{document}

\begin{frontmatter}

\title{A technique for the study of (p,n) reactions with unstable isotopes at energies relevant to astrophysics}

\author[CMU,JINA]{P.~Gastis\corref{cor1}}
\author[CMU,JINA,NSCL]{G.~Perdikakis\corref{cor2}}
\author[ND]{G.P.A.~Berg}
\author[ND,JINA]{A.C.~Dombos}
\author[CMU,JINA]{A.~Estrade}
\author[CMU,JINA]{A.~Falduto}
\author[CMU]{M.~Horoi}
\author[MSU_C,NSCL,JINA]{S.N.~Liddick}
\author[MSU]{S.~Lipschutz}
\author[NSCL,JINA]{S.~Lyons\fnref{label2}}
\author[NSCL,JINA]{F.~Montes}
\author[MSU,JINA]{A.~Palmisano}
\author[NSCL,JINA]{J.~Pereira}
\author[NSCL,JINA]{J.S.~Randhawa}
\author[MSU]{T.~Redpath}
\author[CMU,NSCL]{M.~Redshaw}
\author[MSU]{J.~Schmitt}
\author[MSU]{J.R.~Sheehan}
\author[NSCL,JINA]{M.K.~Smith}
\author[CMU,JINA]{P.~Tsintari}
\author[NSCL,FRIB]{A.C.C.~Villari}
\author[CMU,JINA]{K.~Wang}
\author[MSU,JINA,NSCL]{R.G.T.~Zegers}

\address[CMU]{Department of Physics, Central Michigan University, Mt. Pleasant, MI 48859, USA}
\address[MSU]{Department of Physics $\&$ Astronomy, Michigan State University, East Lansing, MI 48824, USA}
\address[ND]{Department of Physics, University of Notre Dame, Notre Dame, IN 46556, USA}

\address[MSU_C]{Department of Chemistry, Michigan State University, East Lansing, MI 48824, USA}

\address [JINA] {Joint Institute for Nuclear Astrophysics: Center for the Evolution of the Elements,
Michigan State University, East Lansing, MI 48824, USA}

\address[NSCL]{National Superconducting Cyclotron Laboratory (NSCL), Michigan State University, East Lansing, MI 48824, USA}

\address[FRIB]{Facility for Rare Isotope Beams (FRIB), Michigan State University, East Lansing, MI 48824, USA}

\cortext[cor1]{gasti1p@cmich.edu}
\cortext[cor2]{perdi1g@cmich.edu}
\fntext[label2]{Present address: Los Alamos National Laboratory, Los Alamos, NM 87545, USA}

\fntext[label2]{Present address: Pacific Northwest National Laboratory, Richland, WA 99354, USA}

\begin{abstract}
We have developed and tested an experimental technique for the measurement of low-energy (p,n) reactions in inverse kinematics relevant to nuclear astrophysics. The proposed setup is located at the ReA3 facility at the National Superconducting Cyclotron Laboratory. In the current approach, we operate the beam-transport line in ReA3 as a recoil separator while tagging the outgoing neutrons from the (p,n) reactions with the low-energy neutron detector array (LENDA). The developed technique was verified by using the $^{40}$Ar(p,n)$^{40}$K reaction as a probe. The results of the proof-of-principle experiment with the $^{40}$Ar beam show that cross-section measurements within an uncertainty of $\sim$25\% are feasible with count rates up to 7 counts/mb/pnA/s. In this article, we give a detailed description of the experimental setup, and present the analysis method and results from the test experiment. Future plans on using the technique in experiments with the separator for capture reactions (SECAR) that is currently being commissioned are also discussed.

\end{abstract}


\end{frontmatter}



\section{Introduction} \label{sec_intro}

Neutron-induced reactions on stable and unstable nuclei play an important role in the synthesis of heavy elements in our Universe~\cite{Cameron1957,Burbidge:1957}. The (n,p) reactions, particularly, are critical to nucleosynthesis in slightly proton-rich conditions that may be realized in core-collapse supernovae. Recent studies show that the $^{56}$Ni(n,p)$^{56}$Co and $^{64}$Ge(n,p)$^{64}$Ga reactions regulate the efficiency of the so-called neutrino-p process ($\nu$p-process)~\cite{Frohlich:2006,Wanajo:2010,Nishimura:2019}, which is responsible for the formation of elements between nickel (Ni) and tin (Sn) in type II supernovae. To calculate the reaction rates of such critical reactions and predict the elemental abundances produced by this type of nucleosynthesis, we currently rely on theoretical models as very limited -- if any -- experimental data exist for neutron-induced reactions away from stability (e.g., see JINA REACLIB database~\cite{Cyburt_2010}).

In this article, we present a newly developed technique and the associated experimental setup that can be used for the study of (n,p) reactions via measuring the reverse (p,n) reactions in inverse kinematics. By providing experimental data for the reverse (p,n) reactions, significant constrains can be applied on the associated theoretical models employed for the calculation of reaction rates~\cite{Gastis2020a}.
The minimum pieces of equipment needed are a stable hydrogen target, a neutron detector, and a magnetic separator. By taking advantage of the radioactive beams of the Facility for Rare Isotope Beams (FRIB), we will be able to perform measurements with nuclei away from stability using the proposed technique. This includes measurements of critical reactions for the $\nu$p-process, such as the $^{56}$Co(p,n)$^{56}$Ni -- the inverse of $^{56}$Ni(n,p)$^{56}$Co --, or other (p,n) reactions on neutron-deficient nuclei relevant to nucleosynthetic mechanisms such as the p-process~\cite{Truran:1997,Frank:1961}. Our approach can constitute a standard method of studying (n,p) reactions relevant to nucleosynthesis that cannot --yet-- be measured differently.

The direct measurement of neutron induced reactions on short-lived nuclei, constitutes an experimental challenge for physicists~\cite{Larsen:2019}.
Low-energy nuclear reactions are typically measured by using fast moving beams and fixed targets. A wide variety of experimental techniques have been developed to carry out such type of experiments efficiently. 
 For the measurement of (n,p) -- or (n,$\gamma$) -- reactions on short-lived nuclei, however, the situation is more complex. In that case, a measurement using a neutron beam necessitates the construction of a radioactive target. While this is feasible under certain circumstances, it is quite restrictive and can only be used for long-lived isotopes~\cite{Wilhelmy2002}. 
 
 Another approach for the study of such challenging reactions is the measurements in inverse kinematics, using radioactive beams. In that case, however, a neutron target is required. Conceptual ideas on the construction of a neutron target have been reported in literature over the last few years~\cite{Reifarth:2017,Reifarth:2013}. According to this approach, an area with a high neutron density can be created by using a neutron spallation source. The neutrons produced from the source slow down in a moderator, and enter a beam pipe where they serve as a target for the incoming radioactive beam. To balance the low density of such a neutron target, the radioactive beam must be circulated in a storage ring. In this way, the number of beam ions that pass through the neutron gas in the unit of time is boosted significantly, enabling the collection of adequate statistics in a reasonable time frame. While this concept is very promising for future experiments, many challenges need to be addressed before it is ready to be used. 

As mentioned in a previous paragraph, in this study we follow an indirect path for the study of (n,p) reactions by performing measurements on the reverse (p,n) reactions using a recoil separator. Radiative capture reactions such as (p,$\gamma$) and ($\alpha$,$\gamma$) at astrophysically relevant energies ($<$10~MeV/nucleon), have been measured using recoil separators since the mid-80's~\cite{Hahn:1987}. The operational concept for the usage of recoil separators is as follows. After the target (typically hydrogen or helium), the recoil nuclei are separated from the unreacted beam using electromagnetic elements. The recoils are then measured at the focal plane of the separator/spectrometer using a counter detector~\cite{Ruiz:2014}. For the case of capture reactions, an achromatic magnetic separator can resolve the products at the focal plane based on their mass-to-charge ratio (m/q). While the method seems straightforward, adequate suppression of the unreacted beam is not always possible, as the resolving power of the system depends on the mass of the involved nuclei. For measurements with heavy nuclei (A$>$30), gamma detectors are usually employed for tagging the secondary gamma-rays from the capture reactions and improving the signal-to-background ratio~\cite{Hutcheon:2008}.

In the case of (p,n) reactions, the beam ions and the reaction products have almost the same m/q ratio as their mass difference is practically  negligible. To separate the two species using electromagnetic elements, one exclusively relies on their difference in momentum. The separation of ions based on their momentum-to-charge (p/q) ratio is achieved by using dipole magnets (spectrometers) and/or velocity filters. In any case, this velocity dependence of the dispersion makes the measurement of (p,n) reactions difficult. This happens because the desired lateral separation of the ions at the dispersive focal plane (where the slits are usually located), depends on the Q value of the reaction and the energy spread of the unreacted beam. The use of windowless gas targets can minimize the energy straggling of the beam, but it is not always possible to eliminate the unavoidable low-energy tails. Therefore, a neutron detector system for tagging the outgoing neutrons from the (p,n) reactions is needed to further separate the reaction products.

It is important to mention at this point that for measurements of (p,n) reactions on light nuclei, i.e. carbon or oxygen, the usage of a recoil separator may not be necessary. Depending on the energy and mass regime, a tracking proportional chamber could be a viable alternative~\cite{Ishiyama:2003}. For heavy nuclei, however, the situation is rather difficult, as it is impossible to reach the desirable resolution to adequately separate the reaction products from the unreacted beam.

In the setup we present in this article, the suppression of the unreacted beam is achieved by operating a beamline section of the ReA3 reaccelerator of NSCL as a recoil separator (for details see section~\ref{sec_setup}), while tagging the neutrons from the (p,n) reactions with a neutron detector. In the forthcoming years, the beamline-separator that we used here could be replaced by the separator for capture reactions (SECAR), which is currently under construction at the ReA3 facility~\cite{BERG201887}. In this article, we report on experimental measurements that demonstrate the application of the technique for the setup at ReA3. The technique, however, is applicable to any separator system of a similar operating principle. Moreover, the application of the technique to ($\alpha$,n) reactions is straightforward.

In the following sections we describe the details of the experimental setup that we have developed at the ReA3 facility for the measurement of (p,n) reactions in inverse kinematics. We discuss the details of the first proof-of-principle experiment with a stable $^{40}$Ar beam, and we extract partial cross-sections for the $^{40}$Ar(p,n)$^{40}$K reaction. The analysis of the experimental data is discussed step-by-step, and the results are compared with independent cross-section measurements of the same reaction in regular kinematics. Overall, we find an excellent agreement between the test experiment and the reference study, indicating the reliability of the newly developed technique. Methods to further improve the experimental setup and plans for future measurements using SECAR are also discussed.

\section{Description of the Technique and Associated Instrumentation.} \label{sec_setup}

\begin{figure*}
\includegraphics[width=0.95\textwidth]{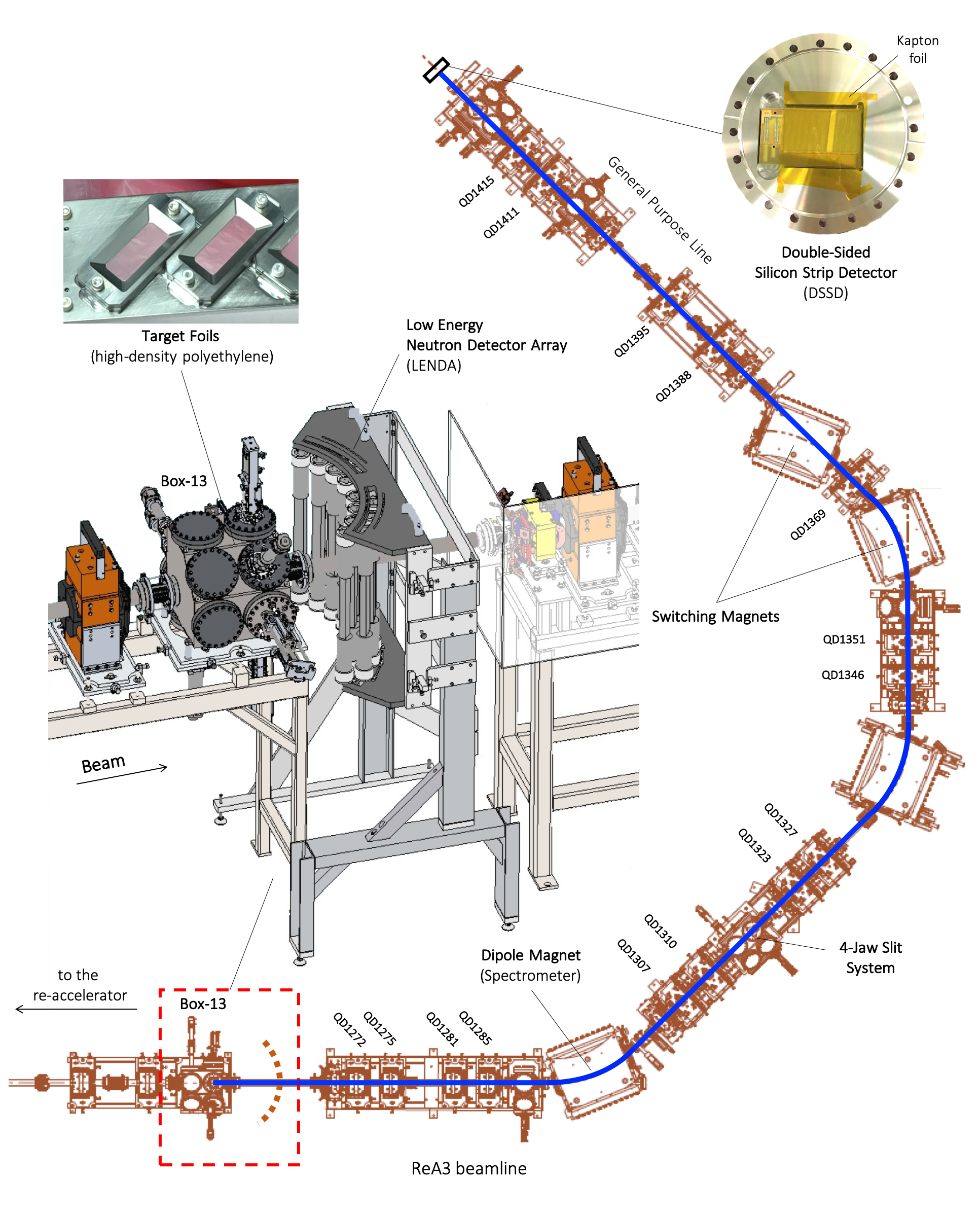}
\caption{ Overview of the experimental setup at the NSCL. The area enclosed in the red rectangle in the figure is the location of the target foils and the LENDA neutron detector. The target foils were placed on the 45 degree drive visible in the blown up drawing of the target area. The vacuum flange housing the double-sided silicon detector is also shown in the photograph in the top-right part of the figure. The photograph shows the  side of the detector setup that the unreacted beam and the reaction products see. The Kapton foil that the products and the unreacted beam have to traverse before getting detected is visible. The solid blue line indicates the path of the nuclei from the target position to the silicon detector. A detailed description of the individual components and the corresponding detection scheme is given in Section \ref{sec_setup}. Mechanical drawings courtesy of the NSCL mechanical design group.}
\label{fig_setup}
\end{figure*}

We present an overview of the experimental setup that was used at the reaccelerator (ReA3) facility of the National Superconducting Cyclotron Laboratory (NSCL) \cite{Kester:2011eia} in Fig. \ref{fig_setup}. We utilized the section of the ReA3 beam-transport line downstream of the last reaccelerator RF cavity. This is the section of beamline that delivers the reaccelerated beam to the user experimental area. The beamline segment we used consists of four dipole magnets (see Fig. \ref{fig_setup}), of which the first two can be operated in an energy-dispersive mode, while the other two are typically used to switch the beam among the various user experimental stations. The experiment made use of the whole beam transport section leading to the general purpose experimental station of the ReA3 experimental area. 

During the measurements, the heavy-ion beam impinged on a solid target made out of high-density polyethylene -- (C$_2$H$_4$)$_n$ -- (HDPE). The target foil was mounted on a target ladder placed inside a vacuum chamber that serves as the diagnostics box 13 of the reaccelerator and is located upstream of the first dipole in Fig. \ref{fig_setup}. We used the low-energy neutron detector array (LENDA) \cite{Perdikakis:2011yf} downstream of that target box to detect the neutrons from the (p,n) reaction. The corresponding heavy reaction products were detected in the vacuum box provided to the users at the end of the general purpose beamline, using a double-sided silicon strip detector (DSSD) of suitable thickness. We tagged the desired reaction events by demanding a coincidence between neutrons detected in LENDA and the heavy ions in the DSSD. The signals from these detectors were use to determine the time-of-flight between the neutrons and the heavy reaction product. A key concept of the experimental setup is the use of the beam-transport line as a recoil separator. The separation between beam ions and reaction products is achieved by operating the first dipole magnet after the target as an energy-dispersive spectrometer.
In this way, the heavy reaction products can be separated from the unreacted beam based on their momentum. We used a 4-jaw slit system at the focal plane of the upstream magnet (see Fig. \ref{fig_setup}) to block the deflected beam ions. While the 4-jaw slits significantly reduced the number of beam ions that could reach the DSSD, it was not possible to entirely eliminate the low-energy tails of the beam.
To further increase the energy separation between the reaction products and the unreacted beam that was transmitted to the experimental end station, we placed an (11.9$\pm$0.6)~$\mu$m thick Kapton foil upstream of the DSSD to increase the energy loss of the ions. The Kapton foil boosted the energy difference ($\Delta$E) between ions with different atomic numbers (Z), facilitating particle identification.  

In the following subsections, we describe the targets, detectors, and the data acquisition system in more detail.

\subsection{Hydrogen Target}\label{sub_sec_target}

For the measurements we used a hydrogen target in the form of a thin HDPE film. The foil was fabricated by dissolving thicker commercially available HDPE films in boiling xylene, according to the procedure described in Ref.~\cite{Tripard:1967}. We determined the target's effective thickness of (3.95$\pm$0.4)~$\mu$m experimentally by measuring the energy loss of a $^{40}$Ar beam in the foil. Considering the nominal density of HDPE (0.95~g/cm$^3$), the corresponding areal density of hydrogen in the target is (3.18$\pm$0.32)$\times$~10$^{19}$~atoms/cm$^2$. By using a thin target foil we were able to achieve a relatively high energy resolution in the measurements (by minimizing the beam energy loss), while maintaining a relatively high density of hydrogen. In order to mitigate any possible out-gassing of the foil inside the target box and maintain the appropriate ultra-high vacuum conditions of the beamline ($\sim$10$^{-8}$ Torr), the target was vacuum baked for approximately 16 hours at 95$^\circ$~C before the experiment.

\subsection{Neutron Detector}\label{sub_sec_lenda}
\begin{figure}[t]
\centering
\includegraphics[width=0.49\textwidth]{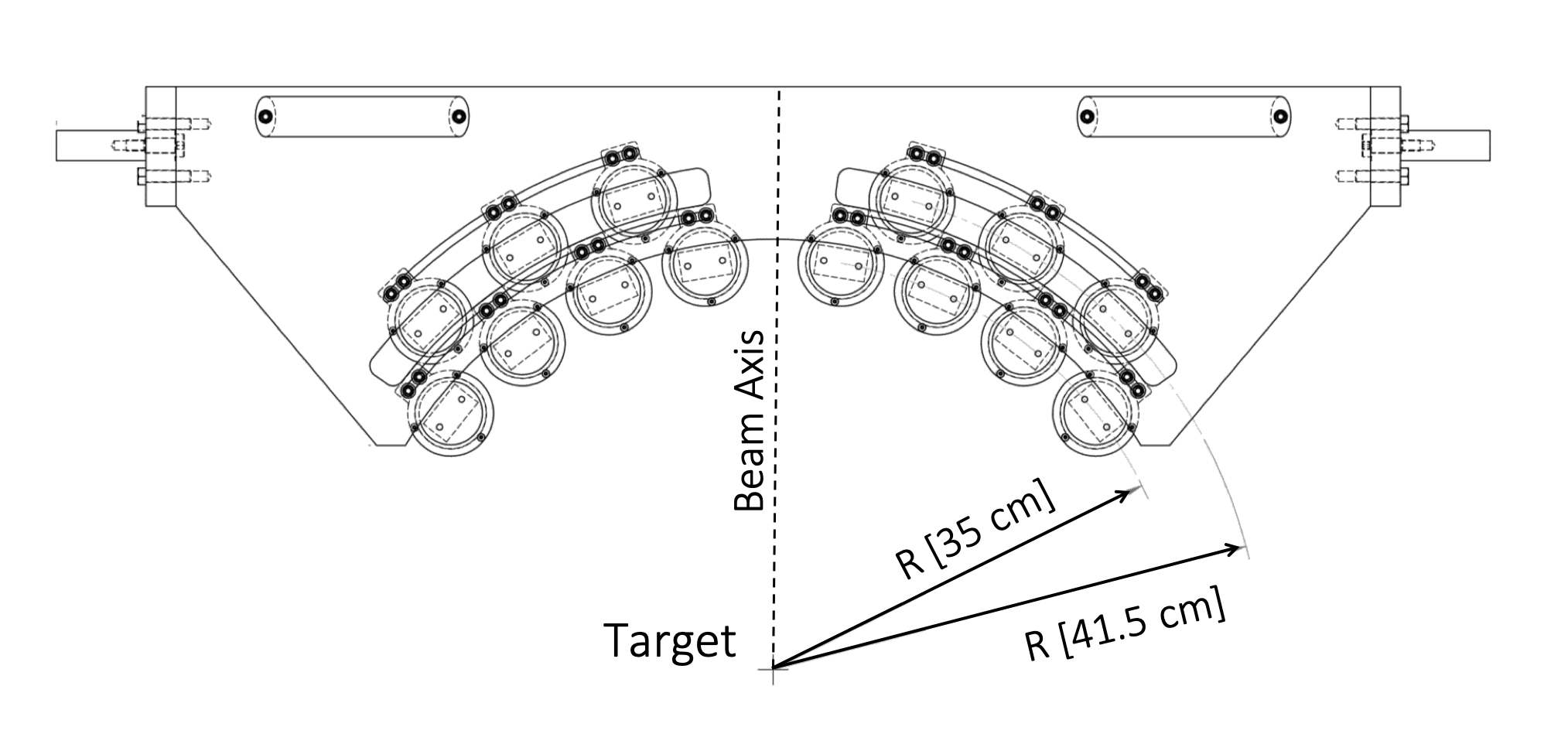}
\caption{ Layout of LENDA in the double-row configuration. This configuration offers near-maximum solid angle coverage for the neutrons from low-energy (p,n) reactions in inverse kinematics.}
\label{fig_lenda}
\end{figure}

We used the LENDA detector array \cite{Perdikakis:2011yf} to register neutrons from the (p,n) reaction. The full detector system consists of twenty-four plastic-scintillator bars made of BC-408 type scintillator. In the current setup, however, we deployed 14 bars arranged in two consecutive layers downstream of the target, as shown in Fig.~\ref{fig_lenda}. By using this configuration, we eliminated the gaps between bars and increased the geometrical solid angle coverage of the array compared to a single layer. A special frame designed to keep the front bars aligned at 35~cm from the target, and the rear bars at 41.5~cm was built for the experiment. Based on this configuration, the angular acceptance of LENDA in the laboratory frame was $ 5^\circ \lesssim \theta _{neut.} \lesssim 52^\circ$. The optimization of LENDA's geometry was done based on Monte-Carlo simulations taking into account the kinematics of low-energy (p,n) reactions in inverse kinematics. Details of these simulations are discussed in Section \ref{sub_sec_solid_ang_analysis}.

\subsection{Double-sided Silicon Strip Detector}\label{sub_sec_dssd}

\begin{figure}[t]
\centering
\includegraphics[width=0.5\textwidth]{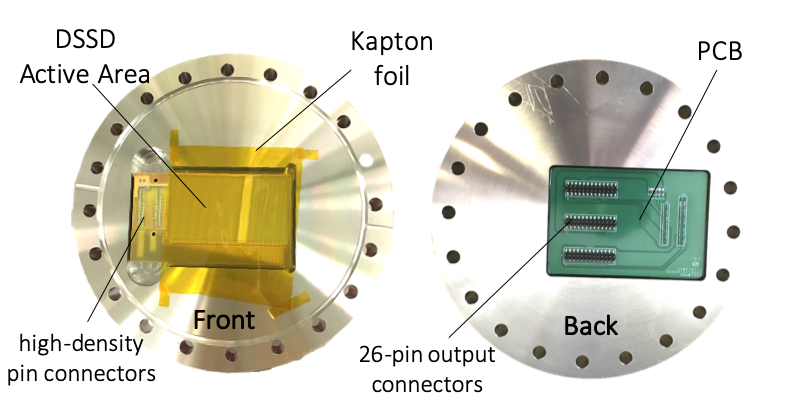}
\caption{Image of the front and the back side of the DSSD flange. The DSSD is connected to the PCB on the back side of the flange through a small rectangular opening located behind the detector. The PCB is ultra-high vacuum compatible and serves both as a feed-through and as an adapter for transferring the signals from the DSSD to the pre-amplifiers. The thin Kapton foil visible in the photo is placed upstream of the DSSD to improve the particle identification (see text for details).
}
\label{fig_dssd}
\end{figure}

The DSSD we used is a standard BB15-type detector from MICRON Semiconductor Ltd. The detector has an active area of (40.3 $\times$ 75.0) mm$^2$ (height $\times$ width) and consists of 64 vertical strips on the junction side (front), and 4 horizontal strips on the Ohmic side (back). The effective thickness of the detector is $\sim$1000~$\mu$m and has an optimum operation bias close to +150~V. Figure~\ref{fig_dssd} shows a picture of the DSSD mounted on a specially designed ConFlat flange. We connected the detector to a custom-made printed circuit board (PCB) that was attached to the air side of the flange. The connection between DSSD and PCB is done through a small vacuum sealed opening on the flange, that allows the detector's output pins to pass-through. The PCB is ultra-high vacuum compatible and serves both as a feed-through and as an adapter to convert the signals from the high-density connectors of the DSSD to the low-density 26-pin connectors at the exit. 

The DSSD signals from the vertical strips of the detector are coupled into groups of two reducing the total number of signal output channels from 68 to 36 (32 vertical and 4 horizontal). The coupling of the strips is done inside the PCB, and the outgoing signals are distributed into three 26-pin connectors on the air side of the board (see the right side of the picture in Fig.~\ref{fig_dssd}). The first two low-density connectors carry the 32 vertical strips (16 strips per connector), and the last one carries the 4 horizontal strips. These modifications reduce the load to the data acquisition system while maintaining a resolution in the x-position of 2.34~mm (the original resolution of the BB15 DSSD based on the width of each strip is 1.17~mm). 

\subsection{Data Acquisition System and Electronics}\label{sub_sec_daq}

\begin{figure*}[t]
\centering
\includegraphics[width=0.99\textwidth]{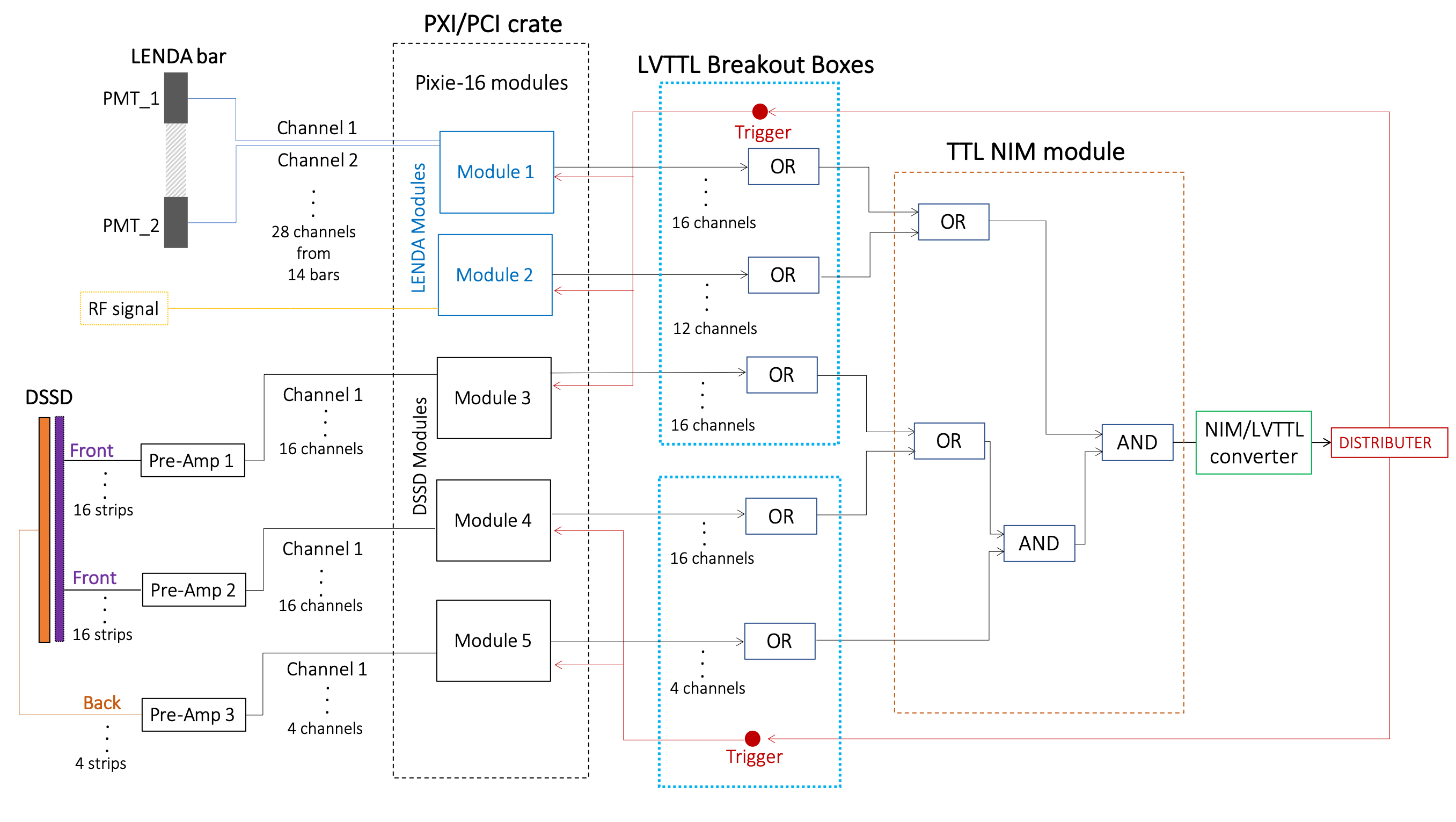}
\caption{Schematic diagram of the data acquisition system. It consists of a total number of 64 signal channels (28 from LENDA and 36 from the DSSD) that are distributed to 5 Pixie-16 cards. The signals are recorded asynchronously and the coincidences between events are registered by the DDAS software code. To enable use of the system in high-rate situations we implemented an optional hardware-based trigger system using two breakout boxes to create an external event validation system. When the external trigger is enabled, the system requires a coincidence between the signals from LENDA and the DSSD to process an event. Since the coincidence condition is processed in hardware before an event is registered the total event rate in DDAS drops significantly.}
\label{fig_electronics}
\end{figure*}

The signals from the DSSD and the photo-multiplier tubes (PMTs) of the LENDA bars, are processed by a digital data acquisition system (DDAS) based on the NSCL DDAS framework~\cite{PROKOP2014163}. A PXI/PCI crate with five XIA Pixie-16 modules is used to collect the 64 input channels, 28 from LENDA (2 channels per bar) and 36 from the DSSD. The Pixie-16 modules are equipped with 250 mega-samples per second analog to digital converters and have 14-bit resolution. This setup is based on a similar NSCL-DDAS configuration that was recently developed and implemented for the LENDA system, which is optimized for higher beam energies~\citep{Lipschutz:2016czw}. 
Figure~\ref{fig_electronics} shows a schematic diagram of the data acquisition system and various aspects of the implemented logic. The signals from the PMTs are directly connected to the Pixie modules, while the DSSD signals are pre-amplified using three MPR-16 pre-amplifiers. An external event validation circuit is available to reduce the amount of data processed by the DDAS when the input rates are too high. The system consists of two LVTTL Breakout Boxes (100~MHz) connected to the Pixie-16 boards, and one TTL logic module. Having the external trigger enabled, the system requires coincidence in the signals from the vertical and horizontal strips of the DSSD, as well as the PMPs of a LENDA bar. While the high input rates do not affect the performance of DDAS, they may have implications on the form of the recorded traces of the signals. In the case of long input signals (e.g. signals from a DSSD), a high input rate will cause overlaps between the traces and increase the complexity of the off-line signal processing. Without the implementation of the external trigger, the highest average input rate on the system before pile-up was of the order of 800~pps. In the measurements described in this work, the coincidence rate on the DAQ never exceeded 400~pps due to the sufficiently effective rejection of the unreacted beam. Hence we did not need to enable the external validation system and all coincidences between LENDA and DSSD events were registered using a software trigger validation. According to that validation scheme, the following logic determines the quality of an event. A signal from a LENDA bar is considered valid when both the PMTs have triggered. Likewise, a signal from the DSSD is valid when at least one vertical (front-side) and one horizontal (back-side) strips have triggered. A valid event must include a valid signal from the DSSD, and at least one valid signal from a LENDA bar within a preset time window. This time window is setup wide enough to cover the time-of-flight (TOF) of the heavy reaction products along the ReA3 beamline, taking into account all relevant delays built into the system via cables and electronics settings.

\subsection{Off-line Analysis Software}\label{sub_sec_analysis}
The data analysis of correlated neutron and heavy residual events was performed by developing appropriate code routines within the R00TLe~\cite{rootle:code} software environment. R00TLe is a software toolkit developed at NSCL to analyze experimental data taken with LENDA using the digital data acquisition system. The core of the software is based on ROOT~\cite{Brun:1997}. The R00TLe toolkit utilizes a number of digital filtering algorithms for extracting timing and energy information from the digitized traces of the LENDA signal. This gives the ability to process the signals off-line and optimize the timing and energy resolution of the system. By integrating the DSSD channels in the LENDA standalone data acquisition system, we were able to use the analysis routines of R00TLe and develop a data analysis tool optimized for the (p,n) experimental setup. 

To get the timing information of each event, the off-line analysis code analyzes the recorded signal traces from the two detectors. For a high-precision timing, the code uses an optimized cubic algorithm to interpolate the digital constant fraction discriminator signal in the region of the zero crossing~\cite{Lipschutz:2016czw}.
For a LENDA bar, this analysis is performed for each PMT separately, and the average timing $\tau$ of an event is given as:
\begin{equation}
\tau = \frac{1}{2}(t_{B} + t_{T})
\label{eq_lenda_time}
\end{equation}
where t$_{T}$ and t$_{B}$ is the timing of the top and the bottom PMT, respectively. With this method, a time resolution of the order of 400~ps can be achieved for a single LENDA bar. On the other hand, the corresponding timing resolution for the DSSD signals is close to 3.5~ns.

\section{Experimental Validation Using the $^{40}$Ar(p,n)$^{40}$K Reaction.} \label{sec_test_exp_valid}

To test the experimental setup and verify the proposed technique, we performed a stable-beam experiment in ReA3 using the $^{40}$Ar(p,n)$^{40}$K reaction as a test case. The choice of reaction was dictated by constraints in well-developed stable beam availability in ReA3 at the time. In this experiment, we determined the partial cross-section of the $^{40}$Ar(p,n)$^{40}$K reaction for transitions to the second excited state of $^{40}$K, using the newly developed setup and a stable $^{40}$Ar beam. We validated the results via a separate experiment at the Edwards Accelerator of Ohio University (OU), where we measured the total and partial cross-sections of the $^{40}$Ar(p,n)$^{40}$K reaction in normal kinematics, using a proton beam and a $^{40}$Ar gas target. Here will present only the results of the OU measurement that are relevant for the comparison with the data taken at ReA3. The full account of the $^{40}$Ar experimental work at OU has been accepted for publication and can be found in Ref.~\cite{Gastis2020a}.

The ion beam was produced by ionizing residual $^{40}$Ar gas inside the Electron Beam Ion Trap (EBIT). EBIT is part of the ReA3 platform upstream of the linac accelerator and is used as a charge-state breeder for the re-accelerated beams that are typically produced at the FRIB facility via fragmentation. The ionized $^{40}$Ar nuclei from EBIT were then transported into the ReA3 bunching system that consists of a Multi-Harmonic Buncher (MHB) and a Radio Frequency Quadrupole (RFQ). This system bunches and focuses the beam ions before injecting them into the re-accelerator. The RFQ also accelerates the ions to an initial kinetic energy of approximately 600~keV/u. The main acceleration takes place downstream of the RFQ in the superconducting linac of ReA3 that is comprised of fifteen Quarter-Wave Resonator (QWR) cavities for acceleration and eight solenoids for beam focusing between the resonators~\cite{Kester:2011eia}. 

For the purposes of our test, the $^{40}$Ar$^{14+}$ beam was accelerated to a final energy of 3.52~MeV/nucleon. Considering the stopping power of $^{40}$Ar in HDPE \cite{Ziegler:125402} and the energy loss of the beam inside the target, the mean reaction energy during the measurement was $(3.42\pm0.12)$~MeV/nucleon, which corresponds to $(3.36\pm0.12)$~MeV in the center-of-mass system. The energy uncertainty of $\pm$3.5\% that we report here is extracted by taking into account the energy loss of the beam ions in the target foil, and the energy spread of the incoming beam.

\begin{figure}
\centering
\includegraphics[width=0.48\textwidth]{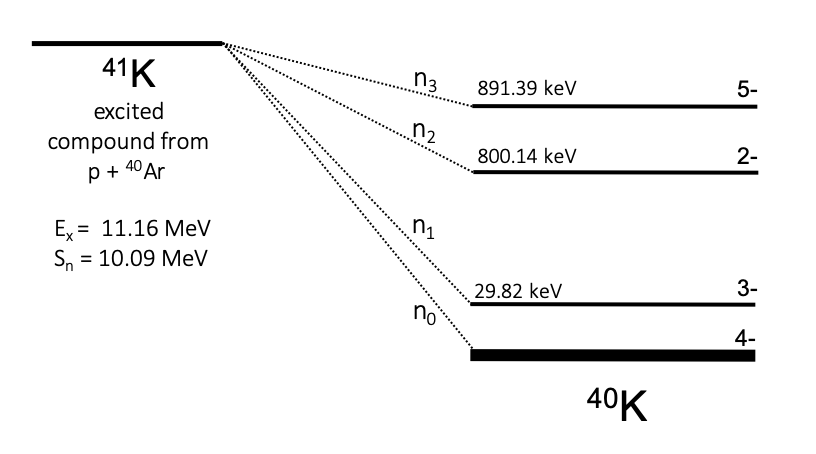}
\caption{Level scheme of the $^{40}$Ar(p,n)$^{40}$K reaction for a compound excitation energy of 11.16 MeV. This corresponds to an incident energy of 3.36~MeV in the center-of-mass system. In this energy regime, the residual $^{40}$K nuclei are allowed to be in any of the first three excited states. }
\label{fig_levels}
\end{figure}

Fig.~\ref{fig_levels}, shows the energy levels of the residual $^{40}$K nuclei that were energetically accessible during this measurement. For an incident energy of 3.42 MeV/nucleon, the $^{40}$K nuclei are allowed to be in any of the first three excited states, and depending on the level of excitation, the corresponding exit channel is different. We refer to these channels as $^{40}$Ar(p,n$_x$)$^{40}$K (x=0,1,2,3). The emittance of the $^{40}$K ions from the different channels varies significantly and this has implications on the transmission of the products through the beam line. The transmission of the reaction products is a critical parameter for determining the reaction cross-section and its effect on the measurements will be described in more detail below.

During the experiment, the LENDA detector and the DAQ system were at a large distance from each other. The LENDA detector close to the target position, and the DAQ close to the DSSD. To connect the two systems, we used 40~m coaxial cables. As a result, neutron signals in the experiment were delayed by $\sim$270~ns, causing a time shift between the actual TOF of $^{40}$K and the registered time of flight difference between LENDA and the DSSD. For the rest of the article we will refer to the time difference between the two detectors as ``TOF." Taking all the timing delays of the test experiment into account, the timing window for the validation of coincidences between LENDA and the DSSD was set to 2~$\mu$s.

In this experiment, we collected two sets of data, with two different ion-optics settings. We refer to the two data sets as `run-1' and `run-2'. The different settings were applied to the section of the beam line that we used as a recoil separator, i.e. between the target position in box 13, and the DSSD at the end of the general purpose beam line. The duration of the two runs was 573~min and 646~min, respectively. By varying the ion optics in these two runs we were able to use the partial cross-section data to explore the optical properties of the ReA3 beam line in the off-line analysis. Using the partial cross-section along with diagnostic measurements of the unreacted beam transmission, we were able to characterize the transmission of ions through the beamline and develop an accurate beam dynamics model. The simulations based on this modeling allow us to reliably extract product transmissions and, consequently, cross-sections in future experiments.

\section{Cross-Section Quantification and Data Analysis}\label{sub_sec_analysis}

In this section, we describe step-by-step the data analysis that we performed in order to extract the partial cross-section for the $^{40}$Ar(p,n$_2$)$^{40}$K channel.

The total cross-section $\sigma$ of the reaction channel that we measured was extracted by using the following equation:
\begin{equation}
 \sigma = \frac{Y_n}{I_b \, N_t \, \epsilon_n \, \omega_n \, T_p \, F_q  }
\label{eq_cross_sec}
\end{equation}
where Y$_n$ is the yield of the reaction products (TOF peak integral), I$_b$ is the total number of beam ions that impinged on the target foil during the measurement, N$_t$ (atoms/cm$^2$) is the areal density of hydrogen atoms in the target, $\epsilon_n$ and $\omega_n$ are the total neutron efficiency and the geometrical solid angle coverage of LENDA, and F$_q$ and T$_p$ are the charge-state fraction and the corresponding transmission of the product nuclei to the DSSD.

\subsection{Number of Correlated Events}\label{sec_yield}

The total number of correlated events in each run was obtained by analysing the corresponding TOF spectra. The total yield Y$_n$ of the reaction products, was calculated as:
\begin{equation}
 Y_n = \sum_{i=1}^{14} Y_i
\label{eq_yield}
\end{equation}
where Y$_i$ is the integral of the TOF peak in the i$^{th}$ LENDA bar. By analyzing the data of each bar separately, we were able to extract information about the neutron angular distribution, and calculate the neutron efficiency of the system more accurately (each LENDA bar has a slightly different intrinsic efficiency).

\subsection{Beam Current Integration}\label{sec_beam_integration}

The total number of beam ions $I_b$ that bombarded the target during each run, was extracted based on the incoming beam current $I$ that was sampled every hour using a Faraday cup. The average value of these measurements was 7.0~pA and 8.1~pA for run-1 and run-2, respectively. 
To account for the fluctuations of the beam current which occur in almost every long measurement, we used the DSSD as a monitor. During the measurements, a small fraction of the unreacted beam was reaching the silicon detector at the focal plane. By integrating the energy peak of the unreacted beam in the off-line analysis, we extracted information about the beam fluctuations as a function of time. To achieve that, we split the total time of each run into $i$ intervals of 5~mins, and for each interval we determined an average beam-rate $I_i$ and a variation $w_i$. The variation of each interval was calculated as $w_i$=$I_i$/$I_{ref}$, where $I_{ref}$ is the average beam-rate of a reference interval. Having extracted the variations $w_i$ of the average current $I$, the total number of beam ions $I_b$ that bombarded the HDPE target during each measurement was calculated as:
\begin{equation}
 I_b = I \, d\tau \, \sum_i w_i
\label{eq_beamrate}
\end{equation}
where $d\tau$ is the width of the time intervals (5~mins).

\begin{figure}
\includegraphics[width=0.47\textwidth]{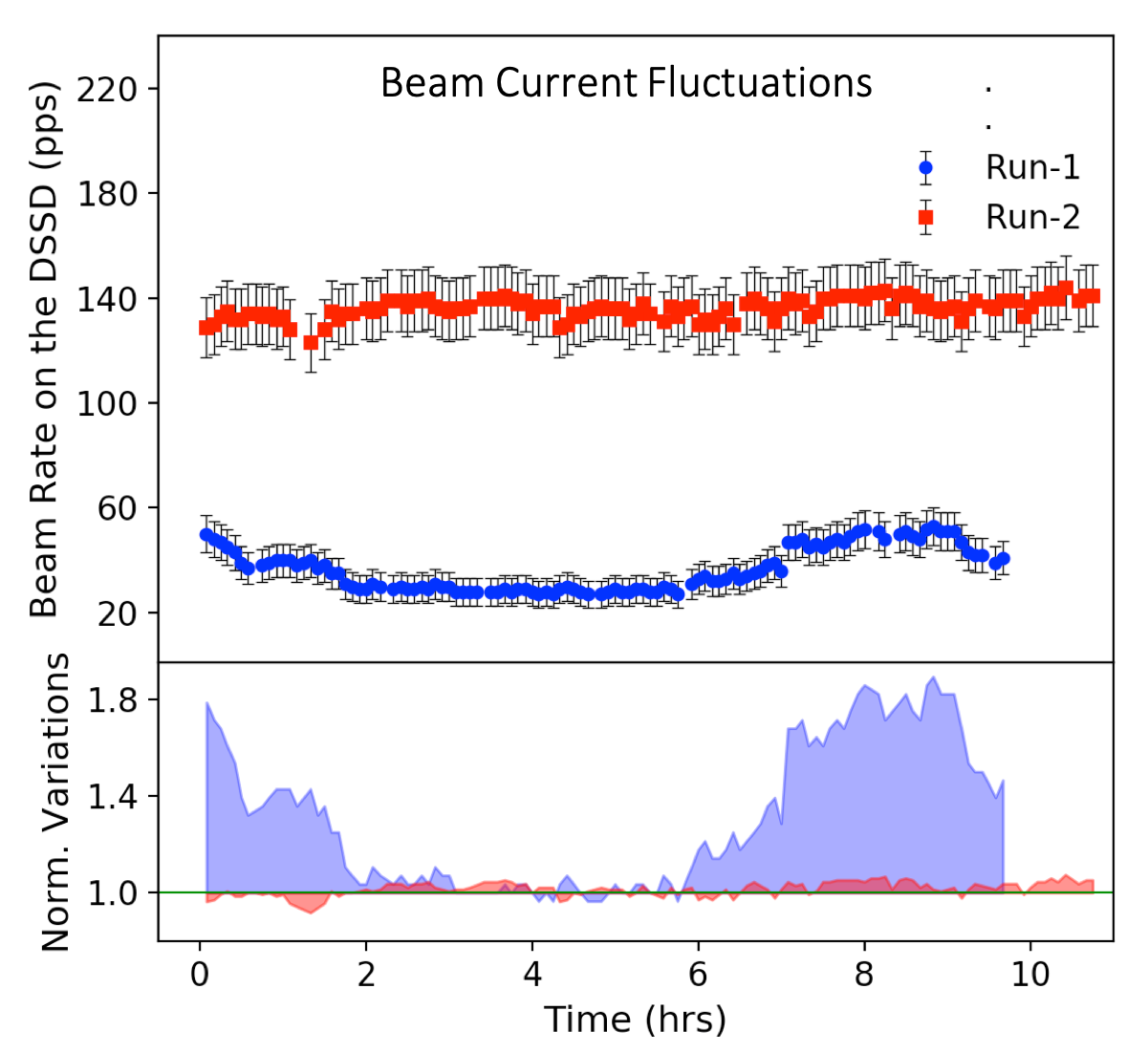}
\caption{Fluctuations of the beam current during the test experiment. (\textbf{Top}) Counting rate of unreacted beam as a function of time during run-1 and run-2. The data points were retrieved by integrating the energy peak of the unreacted beam on the DSSD in off-line analysis. Each point corresponds to the average counting rate in a time interval of 5~mins. (\textbf{Bottom}) Variations of the counting rate relative to a reference value, as a function of time. The red area corresponds to run-1, and the blue area correspond to run-2. In both cases, the reference rate was the one after 4~hours of beam time. To extract the integrated beam current, the normalized variations w$_i$ from this analysis, were implimented in Eq.~\ref{eq_beamrate}.}
\label{fig_beam_rate}
\end{figure}

Figure~\ref{fig_beam_rate} shows the fluctuations of the beam current as a function of time, based on the count-rate of the unreacted beam that was transmitted to the experimental end station. For run-1, we found some smooth variations of the current as a function of time. For run-2, the beam was remarkably stable for the whole period of 11~hours. The relative variations of the beam rates are shown in the lower panel of Fig.~\ref{fig_beam_rate}. The variations of each run are normalized to a reference point in time with negligible fluctuations. In both runs, as a reference point $I_{ref}$, we chose the one at 4~hrs. Taking these data into account, the beam current integral was extracted using Eq.~\ref{eq_beamrate}. The results of these calculations are given in Table~\ref{tab:beam_rate}.

\begin{table}[h]
\caption{Results from beam current integration. The average rates were extracted from the measurements with the Faraday cup. The integrated current I$_b$ was calculated using Eq.~\ref{eq_beamrate}.}
\label{tab:beam_rate}
\begin{tabular*}{\columnwidth}{cccc}
 \rule{0pt}{0.5cm} Run  & Duration & Average Rate & I$_b$ \\ 
   & (min) & (pps) & (atoms) \\ 
 \hline
 \rule{0pt}{0.36cm}
 1 & 573 &  (3.16$\pm$0.32)$\, \times \, $10$^{6}$ & (1.37$\pm$0.14)$\, \times \,$10$^{11}$ \\
 2 & 646 &  (3.63$\pm$0.36)$\, \times \,$10$^{6}$ & (1.43$\pm$0.14)$\, \times \,$10$^{11}$ \\
\end{tabular*}
\end{table}

\subsection{Emittance of Product Nuclei and Beam Dynamics Simulations}\label{sub_sec_dynac_simul}

The counting rate of the reaction products during an experiment can be many orders of magnitude lower compared to the incoming beam rate. This sets vast limitations on the usage of beam diagnostics devices, such as Faraday cups, for the direct measurement of products' transmission (T$_p$) along the various sections of the beamline. As a result, these properties can only be determined through analytical simulations. 

The transmission of the $^{40}$K ions from the $^{40}$Ar(p,n)$^{40}$K reaction during the proof-of-principle experiment was extracted from beam dynamics simulations using the code DYNAC~\cite{Tanke2002DYNACAM}. Specifically, we used a detailed model of the entire ReA3 beamline that has been developed recently and is currently used as a diagnostic tool and for beam optimization~\cite{Wittmer:2014mwa,Yoshimoto:2017exd,Alt2016}. For the purpose of our data analysis, we run simulations using only the part of the transport line between the target position and the DSSD (see Fig.~\ref{fig_setup}). To make the beamline model more realistic and reproduce the system's acceptance more accurately, we added a number of apertures along the beamline that match the shape and size of the actual beam pipes. 

To determine the system's acceptance and the transmission of the ions through the beamline, two important parameters must be specified. The first one is the initial emittance of the incoming ions and the second is the pole-tip fields of the quadrupole magnets during the experiment. The experimental values of the pole-tip fields were determined based on the applied electric current on each quadrupole magnet during the measurement.
The initial emittance of the beam was calculated based on the expected energy and angular spread of the ions, and the size of the beam spot. 

In DYNAC, the initial beam profile can be generated by using the random generators that are implemented in the code (e.g. see Ref.~\cite{dynac_gui}). The generators set the 6D particle coordinates based on the input parameters provided by the user. The 6D coordinates are defined by the longitudinal parameters $\delta$E and $\delta\tau$, that give the relative energy difference and the relative time difference of each particle from a reference point, and the transverse parameters x, y, p$_x$, and p$_y$, that give the positions and directional angles of the particles relative to the horizontal(x) and vertical(y) axes, respectively. These angles are the Cartesian components of the angle $\theta$ relative to the beam axis:
\begin{equation}
 \theta = \sqrt{ p_x^2 + p_y^2}
\label{eq_theta}
\end{equation}
In the used coordinate system, the beam-axis corresponds to the z-axis. Within this context, the initial phase spaces (x,p$_x$), (y,p$_y$), and ($\delta$E, $\delta\tau$) are generated based on Gaussian probability distributions of the form: 
\begin{equation}
 f_G(x) = A\, exp\left[ - \left(\frac{x-\mu_x}{\sqrt{2}\sigma_x}\right)^2 \right]
\label{eq_gauss}
\end{equation}
where A is a normalization constant, and $\mu_x$, $\sigma_x$ are the mean value and standard deviation of the quantity x (x = E, p$_x$, p$_y$, etc). The $\mu$ and $\sigma$ parameters of each quantity in DYNAC are provided by the user.

\begin{figure}
\includegraphics[width=0.5\textwidth]{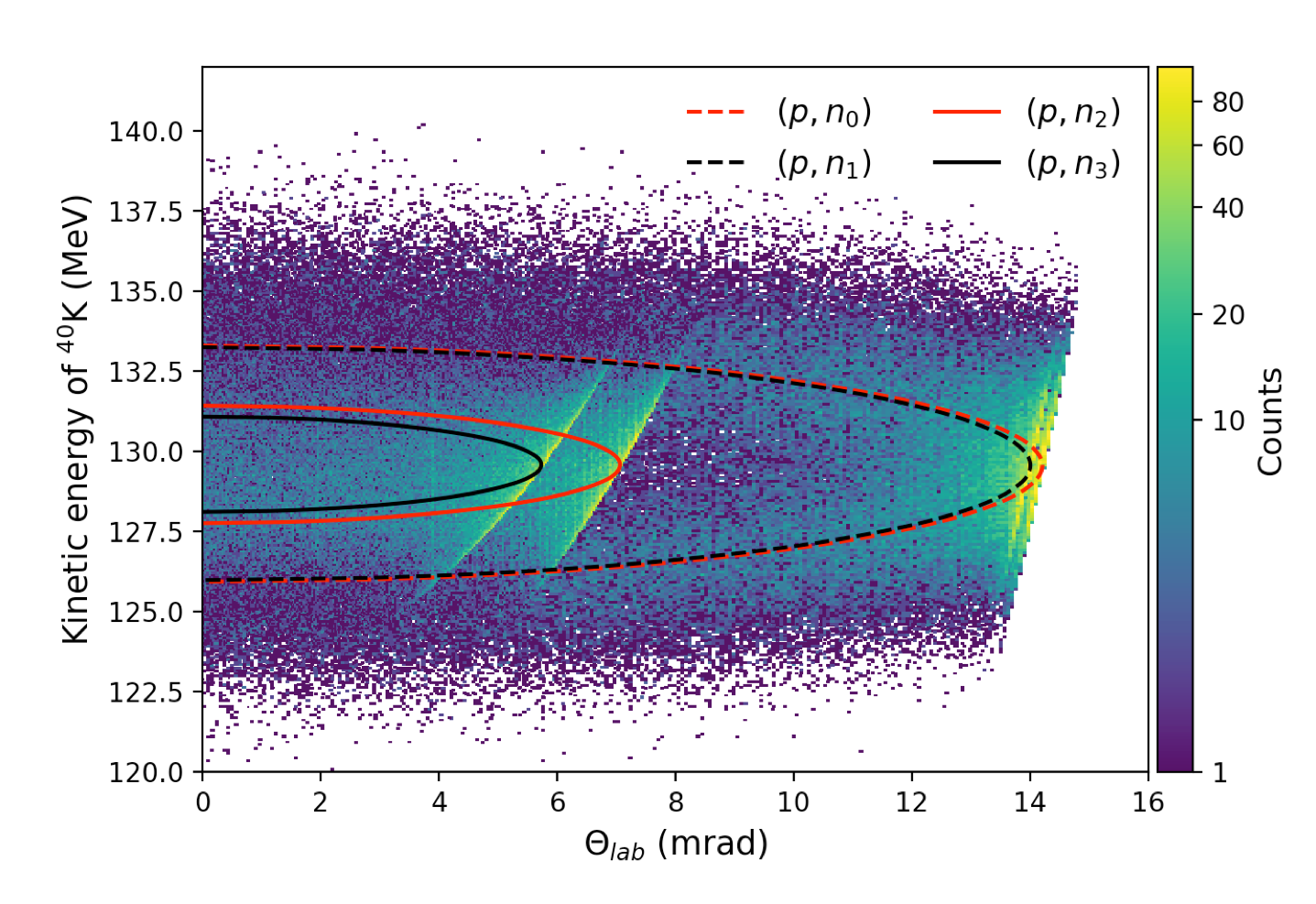}
\caption{Kinematics plot for the $^{40}$Ar(p,n)$^{40}$K reaction. The solid and dashed lines show the kinematic lines of the different exit channels for an incident energy of 3.42~MeV/nucleon, which is the mean reaction energy during the test experiment. This 2D histogram is the result of a Monte-Carlo calculation and shows the energy and angle of the emitted $^{40}$K nuclei that we obtained by randomizing the reaction energy. The reaction energy could take any value between the incoming and exit beam energy (i.e. after the target). The extent of this energy range is visible in the yellow regions in the kinematics plot, and corresponds to the total energy loss of $^{40}$Ar in the HDPE target.}
\label{fig_kinematics}
\end{figure}

To extract realistic angular and energy distributions for the reaction products, we performed Monte-Carlo kinematics calculations, taking into account the interactions of the residual ions with the target atoms, as well as the details of the kinematics of the various exit channels (see Fig.~\ref{fig_levels}). The kinematics plots of the $^{40}$Ar(p,n)$^{40}$K reaction for the various exit channels are presented with the solid and dashed lines in Fig.~\ref{fig_kinematics} (solid and dashed lines). As indicated by these lines, the excitation energy of the emitted $^{40}$K ions affects their energy and angular spread. Additionally, the energy loss of the beam ions inside the target has a strong impact. To investigate this effect, we repeated the kinematics calculations for various incident energies that we randomized based on a Gaussian distribution (see Eq.~\ref{eq_gauss}), where the mean energy E$_0$ of the distribution was the mean reaction energy (3.42~MeV/nucleon),
 and $\sigma_E$ was the corresponding energy spread (47~keV/u). Here, we assume that the full-width-at-half-maximum (FWHM) of this distribution ($\approx$2.355$\sigma_E$) is equal to the energy loss of the $^{40}$Ar beam in the HDPE target. The energy loss was determined experimentally by measuring the difference in the magnetic rigidity (B$\rho$) of the dipole magnets before and after the target foil and also with and without the target foil for the same magnet. The result of the Monte-Carlo calculation is the 2D histogram in Fig.~\ref{fig_kinematics}.

 To obtain the energy and angular characteristics of the emerging $^{40}$K ions after the target foil, the corresponding energy loss of the ions as well as the energy and angular straggling inside the target foil must be taken into account. 
 Considering a 3.95~$\mu$m-thick HDPE target we were able to convert the randomly chosen reaction energy into position inside the target. This was done based on the stopping power models of~\citet{Ziegler:125402}. Having determined the position along the target's length where each random reaction event takes place, the effective thickness of the HDPE layer that each product nucleus has to traverse before exiting was obtained. Taking this into account, the initially assigned angle $\theta$ and energy $E$ of each product were sampled using appropriate Gaussian probability distributions (see Eq.~\ref{eq_gauss}) with the mean values $\mu_\theta$ and $\mu_E$ of these distributions taken from kinematics, and the standard deviations $\sigma_\theta$ and $\sigma_E$ calculated using the SW~\cite{Sigmund:1974,Sigmund:1975} and LS~\cite{Lindhard:1996} models, respectively (see also~\cite{Geissel:2002,Atima}). To take into account the angular spread due to the angular spread of the beam ions, the corresponding dispersion $\sigma_\theta$ was increased accordingly based on the angular straggling of $^{40}$Ar in HDPE.

\subsection{Charge-State Fractions}\label{sub_sec_charge_fractions}

The number of product nuclei that are transported to the DSSD depends on their corresponding charge-state distribution (CSD) after the target. The charge state fractions F$_q$ of the product nuclei are usually determined experimentally by performing CSD measurements with stable beams. For the $^{40}$Ar(p,n)$^{40}$K reaction, this would be done by using a $^{40}$K beam and a HDPE target. In this way the beam intensity would be sufficiently high for determining the associated CSD of $^{40}$K precisely. Such a beam, however, is not available at the ReA facility. Hence, in the present study, no such measurements could be performed in a reasonable time due to the limitations caused by the low intensity and broad angular spread of the $^{40}$K products. Nevertheless, we performed measurements for the $^{40}$Ar beam, which has a fairly similar atomic structure with $^{40}$K.

The charge-state fractions of the $^{40}$K$^{+q}$ ions after the target were determined by using semi-empirical models. Particularly, we used the semi-empirical model of Schiewitz~\cite{Schiwietz:2004}, which was found to give the most accurate results when tested against the experimental data from $^{40}$Ar (see Sec.~\ref{sub_sec_charge_states}). The fractions were extracted based on a Gaussian distribution of which the mean value $\mu$ and standard deviation $\sigma$ were given from the semi-empirical formulas of the model~\cite{Datz:1983}.  For the charge-state fractions extracted with this method, a conservative error of 10\% was estimated based on previous comparisons with experimental data (e.g. see~\cite{Gastis:2015cfa}).

\subsection{Intrinsic Efficiency and Solid Angle of LENDA}\label{sub_sec_solid_ang_analysis}

\begin{figure}
\includegraphics[width=0.98\linewidth]{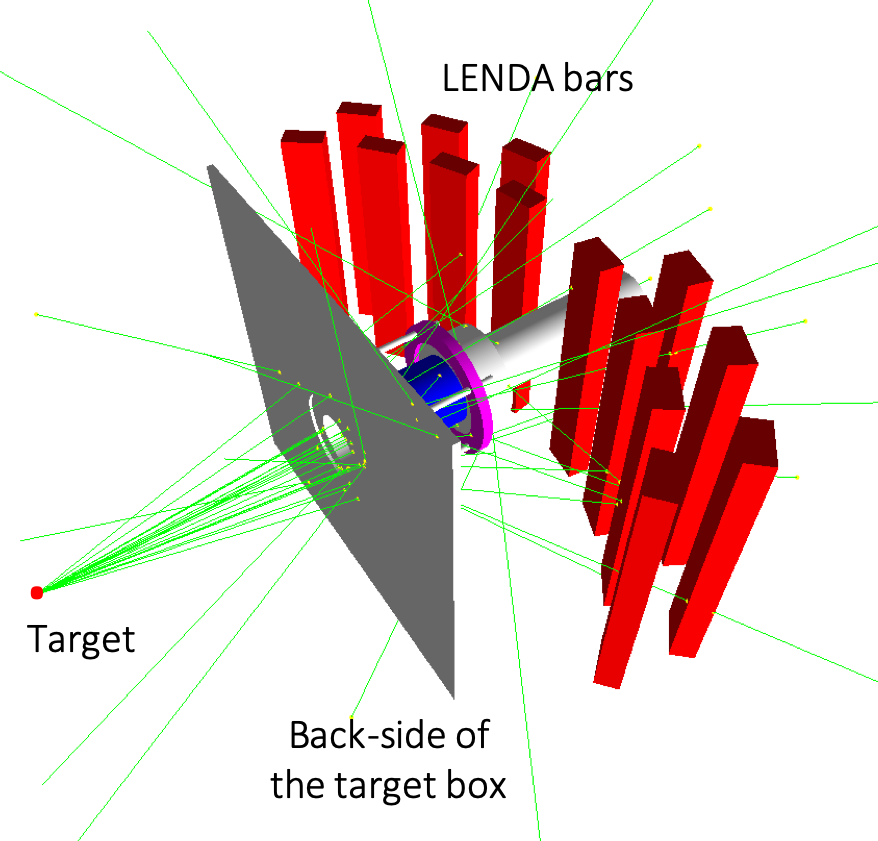}
\caption{Graphical representation of the experimental setup around the target in GEANT4. In this simulation, the solid angle and the intrinsic neutron efficiency of LENDA were determined. The extracted neutron angular distribution was found to be in a good agreement with the experimental data, indicating the adequate accuracy of the simulation (see Sec.~\ref{sub_sec_solid_ang}). }
\label{fig_geant}
\end{figure}

The geometrical solid angle $\omega_n$ and the intrinsic neutron efficiency $\epsilon_n$ of the LENDA system were calculated through a Monte-Carlo simulation with GEANT4~\cite{Agostinelli:2002hh}. Figure~\ref{fig_geant} shows the geometry of LENDA in GEANT4 along with beamline components interposed between the target and the detector. The energy and angular characteristics of the incoming neutrons were generated based on the kinematics of the $^{40}$Ar(p,n$_2$) reaction, taking into account that the shape of the neutron angular distribution in the center-of-mass system is isotropic~\cite{Gastis2020a}. In this simulation, the intrinsic neutron efficiency of each bar was taken into account. 

The intrinsic neutron efficiency of a single LENDA bar depends on the applied light output threshold, which is typically expressed in electron equivalent energy units -- keV$_{ee}$ in this case --. By implementing the experimental threshold level of each bar in the simulation, the intrinsic efficiency curves were obtained. This method of extracting the LENDA efficiency has been successfully used in previous work~\cite{Perdikakis:2011yf,Lipschutz2018} and has been validated against experimental measurements~\cite{Sasano:2012,Sasano:2011}. 

\begin{figure}
\includegraphics[width=0.98\linewidth]{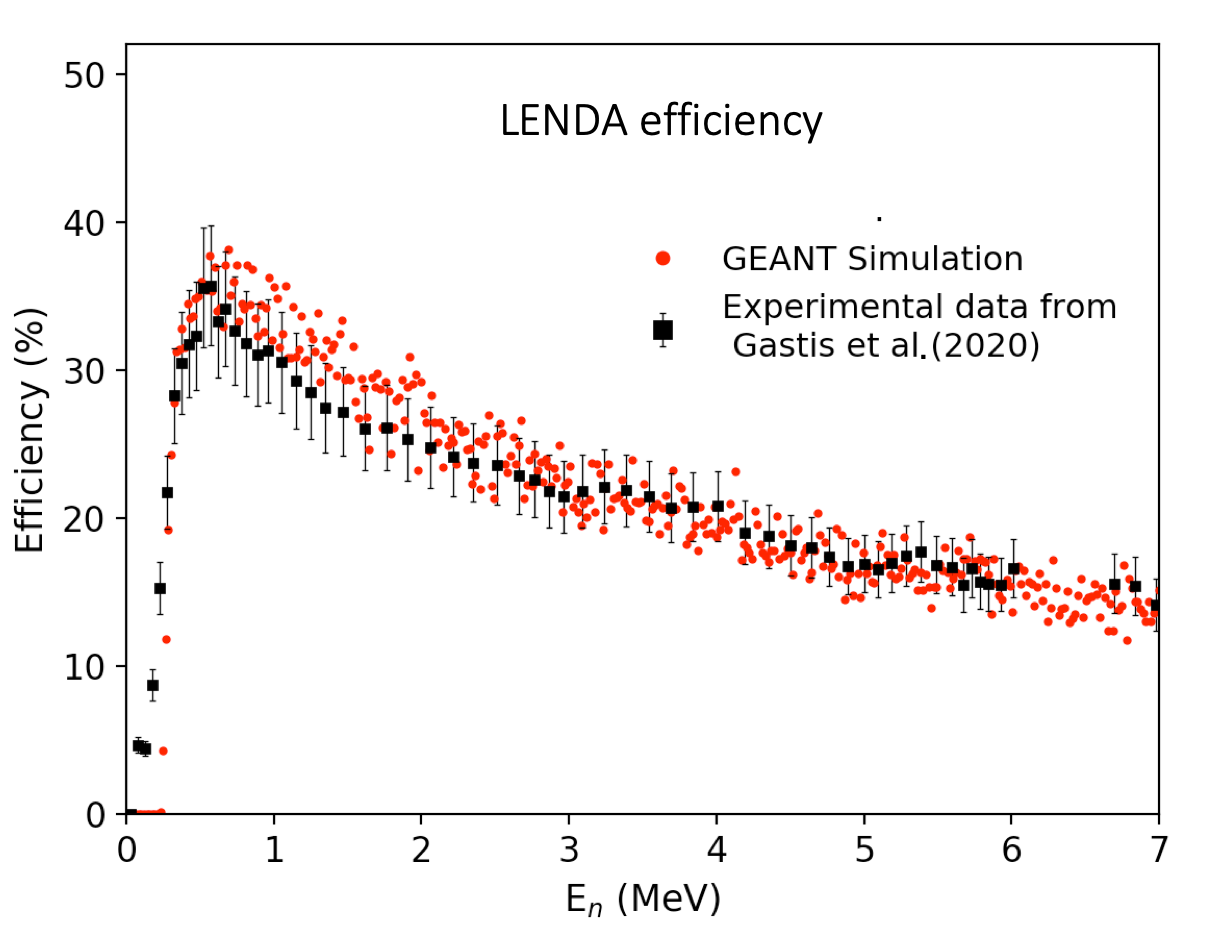}
\caption{Intrinsic efficiency of a single LENDA bar. Experimental data for the efficiency of LENDA were retrieved from~\cite{Gastis2020a}. The theoretical data points were extracted from Monte-Carlo calculations using GEANT4. A scaling factor of 0.92 was applied on the simulated curve, as suggested in~\cite{Lipschutz2018}. In both the experimental and calculated data sets, the light-output threshold was 30 keV$_{ee}$. Within the corresponding uncertainties, the agreement between the two curves is adequate. }
\label{fig_eff}
\end{figure}

It is important to mention, that the calculated intrinsic efficiencies from these simulations had to be scaled. The scaling factor adopted in this study was 0.92 and has been extracted by comparing the simulation to the experimental efficiency of LENDA measured with a $^{252}$Cf fission source~\cite{Lipschutz2018,Perdikakis:2011yf}. To cross-check the accuracy of this scaling, the simulations were tested against experimental data from an independent efficiency measurement~\cite{Gastis2020a}. In that study, the efficiency of a 31~cm long LENDA bar --1~cm longer than the ones used in the current work-- was measured by using a white neutron spectrum from the $^9$Be(d,n) reaction. The comparison between the experimental efficiency and the corresponding GEANT4 simulation for a single bar are presented in Fig.~\ref{fig_eff}. In both data sets, the light-output threshold was set to 30~keV$_{ee}$. The scale-down factor of 0.92 suggested by~\cite{Lipschutz2018} is sufficed to achieve a very good agreement between the two data sets within the experimental uncertainty of $\pm$11\% in Ref.~\cite{Gastis2020a}.

The extracted geometrical solid angle $\omega_n$ and the intrinsic efficiency $\epsilon_n$ of LENDA from these simulations were (54.5$\pm$3.0)\% and (21.9$\pm$2.4)\%, respectively. The results are tabulated in Table~\ref{tab:results_solid_ang}. The error of the solid angle was estimated based on the sensitivity of the results on the geometry of the system and the characteristics of the incoming neutrons, while for the intrinsic efficiency, the experimental uncertainty from Ref.~\cite{Gastis2020a} was adopted. 

The validity of the above GEANT4 simulations was tested by comparing the simulated neutron angular distribution with experimental data and is discussed in Sec.~\ref{sub_sec_solid_ang}.

\begin{table}
\centering
\caption{Geometrical solid angle coverage and intrinsic neutron efficiency of LENDA during the experiment.}
\label{tab:results_solid_ang}
\begin{tabular*}{0.95\columnwidth}{cc}
 \rule{0pt}{0.5cm} 
 Solid angle $\omega_n$ (\%)  & Intrinsic efficiency $\epsilon_n$ (\%) \\ \hline
 \rule{0pt}{0.36cm}
 54.5$\pm$3.0 & 21.9$\pm$2.4  \\
\end{tabular*}
\end{table}

\subsection{Sputtering of Target Material}\label{sub_sec_sputtering}

The loss of hydrogen from the target due to sputtering of target material in this experiment was considered negligible. This is based on the fact that the beam intensity during the measurements was considerably low ($\sim$10$^6$ pps). To check the validity of this assumption, we performed Monte Carlo simulations using the software SRIM~\cite{SRIM}. In these simulations, a high density polyethylene foil of proper thickness and a $^{40}$Ar beam at 140 MeV were considered. The calculated sputtering yield was of the order of 1 atom of H per 100 ions of $^{40}$Ar. The total number of beam ions impinged the target foil during the experiment was $\sim$10$^{11}$ (see Table~\ref{tab:beam_rate}, Sec.~\ref{sec_beam_integration}). Based on the above calculation, the expected number of hydrogen atoms sputtered out of the target in a period of $\sim$24 h is close to 10$^{9}$ atoms. This number is negligible compared to the total number of hydrogen atoms in the target foil, which is of the order of 10$^{19}$ atoms/cm$^2$ (or $\sim$10$^{17}$ atoms/mm$^2$).

\section{Results from the $^{40}$Ar Experiment}\label{sec_results}

\subsection{Particle Identification}
\begin{figure}
\includegraphics[width=0.48\textwidth]{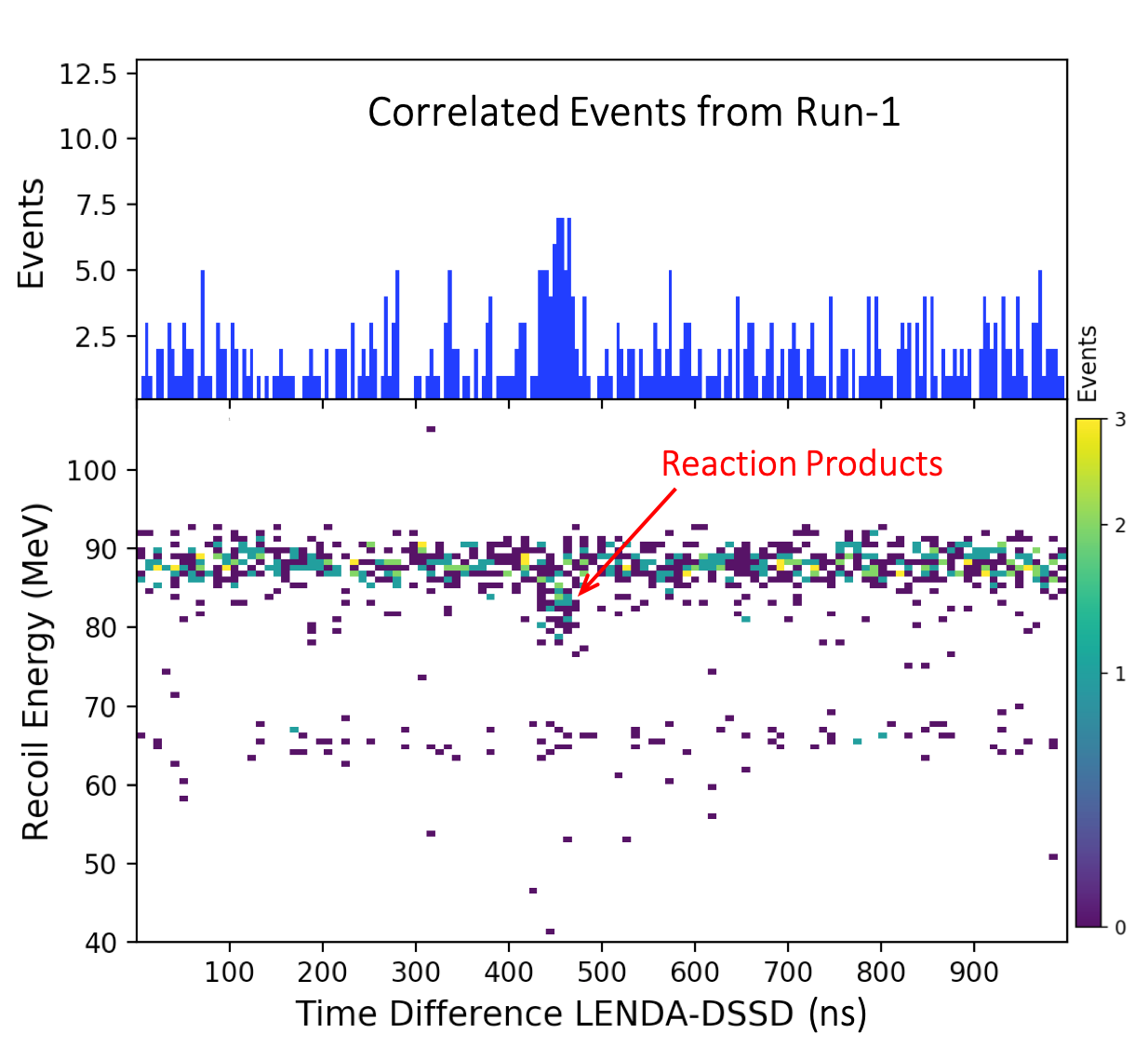}
\caption{Correlated events from run-1. These events satisfy the basic condition of coincidence between the two detectors within a time window of 2~$\mu$s. (\textbf{Bottom}) 2D histogram of the energy deposited in the DSSD, versus the time difference between the signals of the two detectors. This time difference is directly proportional to the TOF of the ions along the beamline. The events from $^{40}$K and neutrons, appear as a blob with specific timing properties. The rest of the events are random coincidences between the unreacted beam ions and the background signals of LENDA. (\textbf{Top}) projection of the 2D histogram on the time axis. A well defined peak appears in the middle of the spectrum that corresponds to the correlated events from the reaction products. This peak was used for extracting the reaction yield, and subsequently, the reaction cross-section.}
\label{fig_run1_corr}
\end{figure}

\begin{figure}
\includegraphics[width=0.48\textwidth]{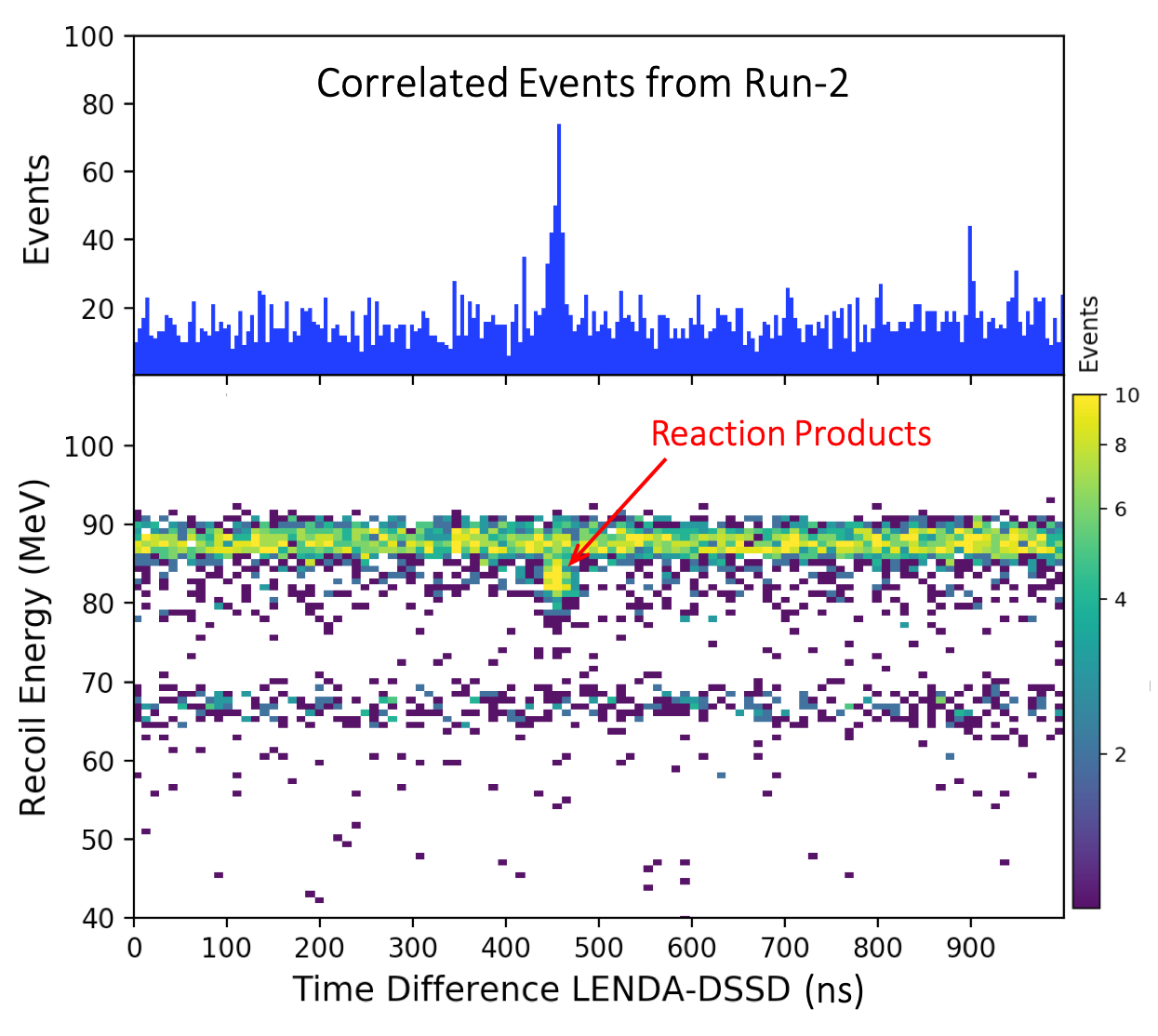}
\caption{Correlated events from run-2. A detailed description of the plots is given in Fig.~\ref{fig_run1_corr}.}
\label{fig_run2_corr}
\end{figure}

In Figures \ref{fig_run1_corr} \& \ref{fig_run2_corr}, we present the energy and timing characteristics of the events that satisfied the coincidence condition between neutron and heavy residual during the two runs of the test experiment. The bottom panels show the energy deposited in the DSSD versus the time difference between the signals from LENDA and the DSSD. The correlated events associated with the products from the $^{40}$Ar(p,n)$^{40}$K reaction appear as a blob at recoil energy of $\sim$84~MeV, that is in agreement with the expected energy of the $^{40}$K ions after passing through the Kapton foil. The time stamp of the products corresponded to about 450~ns. This time is consistent with the TOF of $^{40}$K for mean energy of $\sim$124~MeV, after taking into account the effects of the long signal cables. Cumulative TOF spectra from run-1 and run-2 are shown in the upper panels of Fig.~\ref{fig_run1_corr} and \ref{fig_run2_corr}. These spectra were obtained by projecting the 2D histograms of the lower panels on the time axis (horizontal axis). To reduce the number of background events, we considered only the events in the range 75~MeV $<$ E$_{recoil}$ $<$ 90~MeV. This condition rejects the events with recoil energies outside the expected energy range for the $^{40}$K ions.

\begin{figure}
\includegraphics[width=0.48\textwidth]{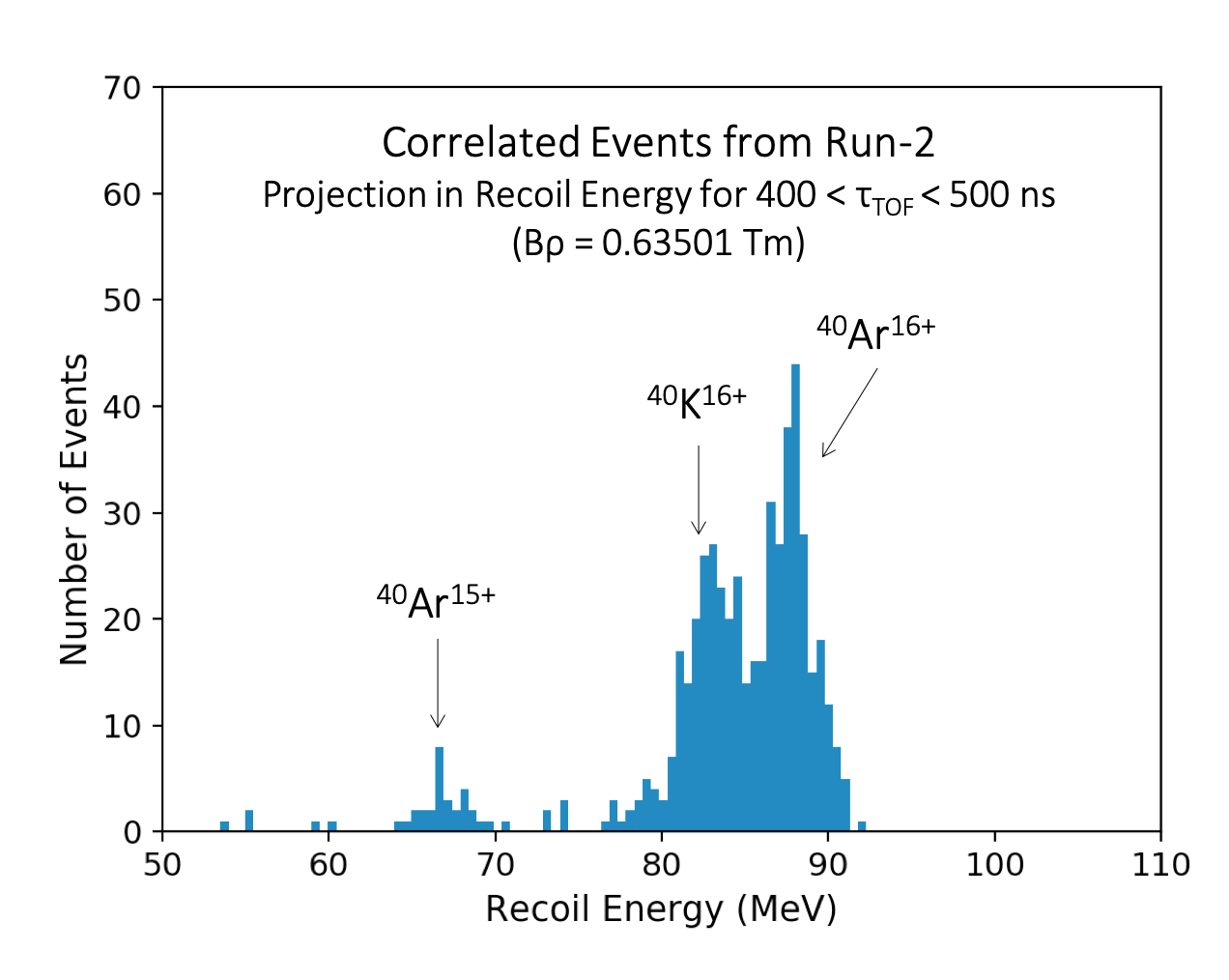}
\caption{Energy spectrum of heavy ions on the DSSD. These events corresponds to time-of-flight values between 400 and 500~ns (see Fig.~\ref{fig_run2_corr}). The energy peak of the reaction products appears between the two peaks corresponding to the 15+ and 16+ charge-states of the unreacted beam. The energies that these charge-states appear are consistent with the B$\rho$ settings of the measurement and the energy loss of the beam in the HDPE target and the Kapton foil upstream of the DSSD.  }
\label{fig_rec_energy_run2}
\end{figure}

All the background events in the spectra (see Figs.~\ref{fig_run1_corr} \& \ref{fig_run2_corr}) are coming from random coincidences between the $^{40}$Ar ions on the DSSD, and the background events on LENDA (cosmic rays etc). Since there is no timing correlation between two random particles that hit the detectors, the time stamp of these events was uniformly distributed within the coincidence window 0 -- 2~$\mu$s. As a result, the random coincidences appear as a linear background in the TOF spectra. For each TOF peak, therefore, the corresponding reaction yield was obtained after subtracting a linear background. Likewise, in the 2D spectra (E vs TOF), the background events appear as a straight horizontal line. As can be seen in the 2D histograms of Figs.~\ref{fig_run1_corr} \&~\ref{fig_run2_corr}, there are two horizontal lines of background events. Both lines are random coincidences associated with the unreacted beam, but each line corresponds to a different charge-state of the beam ions. For example, in Fig.~\ref{fig_rec_energy_run2}, we show a recoil energy spectrum from run-2. The plot was extracted by projecting the corresponding 2D histogram of Fig.~\ref{fig_run2_corr} (E vs TOF) on the energy axis. The displayed events are gated on TOF in order to reduce the unwanted background. The position of the peaks matches the expected energies of the unreacted beam charge-states for that particular B$\rho$. As we see, in run-2 we observed the 15+ and 16+ charge-states of the beam. In run-1, due to the different optics settings the corresponding charge-states were the 15+ and 14+.

In the above analysis, the TOF peaks were exclusively assigned to events coming from neutrons and $^{40}$K ions from the $^{40}$Ar(p,n$_2$) channel. In principle, some of these events could involve photons instead of neutrons, as there is a number of characteristic gamma-rays (de-excitation of $^{40}$K, see Fig.~\ref{fig_levels}) that are also emitted in coincidence with the heavy products. This could add a number of events in the TOF peaks that must be analyzed separately, as they correspond to a different intrinsic efficiency and solid angle. However, the geometrical solid angle coverage of LENDA for any emitted photons is less than 5\%. This happens because the secondary photons are emitted uniformly in 4$\pi$, and are not boosted at forward angles as it happens with the neutrons. Furthermore, the intrinsic efficiency of LENDA for incoming photons is of the order of 1\%. As a result, any contribution from events due to secondary photons was considered negligible for this analysis. 

\begin{figure}[h]
\includegraphics[width=0.48\textwidth]{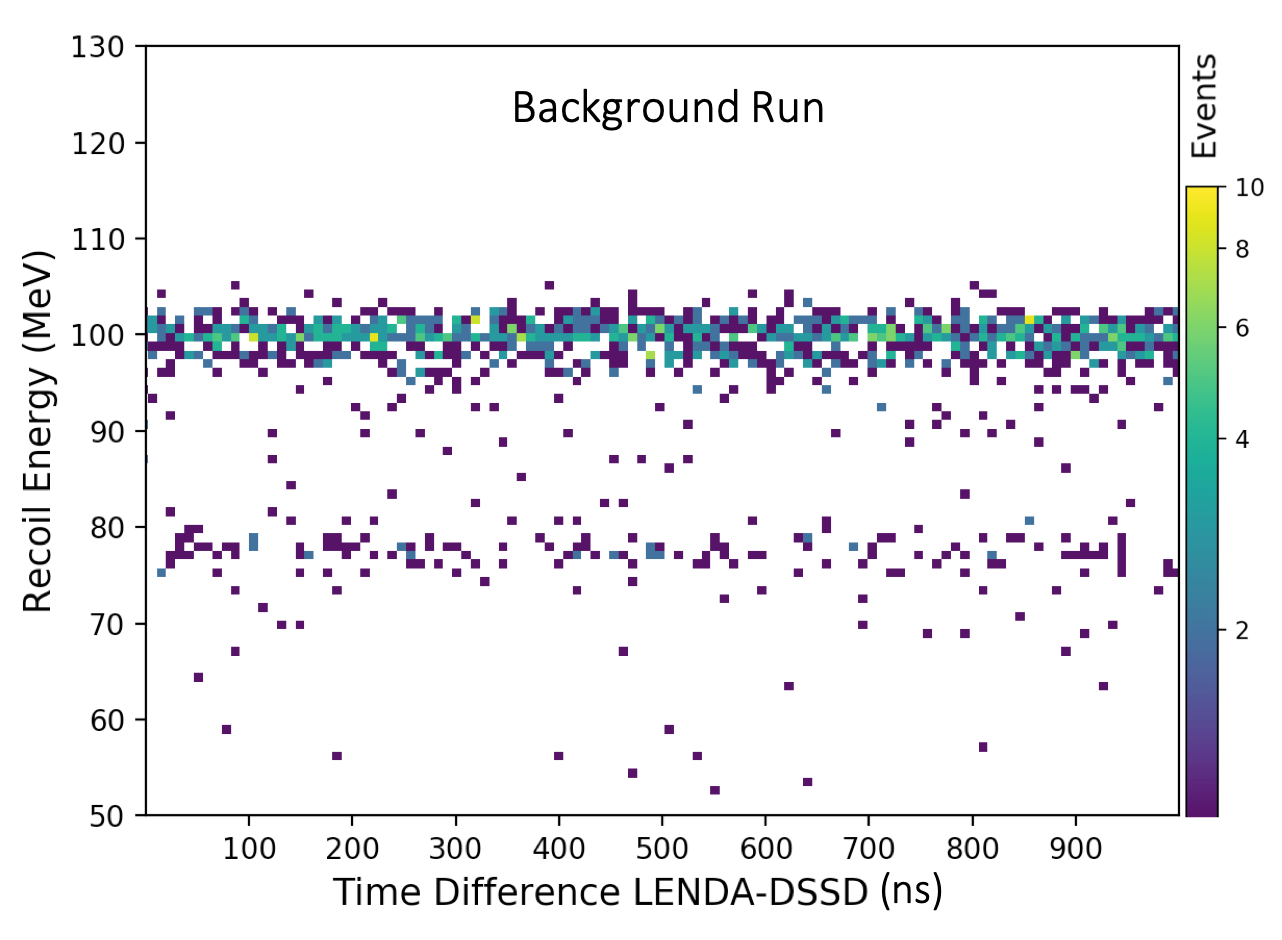}
\caption{Correlated events from the background run. A detailed description of the quantities presented in this 2D histogram are given in Fig.~\ref{fig_run1_corr}. During the background run, a thin carbon foil was irradiated for about 5~hours with the $^{40}$Ar beam. No events with particular time stamp were observed, indicating the absence of background events from carbon in the TOF spectra that could influence the measured yield of $^{40}$K. }
\label{fig_bgr_corr}
\end{figure}

Finally, to investigate whether the peak of correlated events may include any parasitic products from the interactions of the beam with the carbon nuclei in HDPE, we performed a 5-hour background measurement using a thin carbon foil (100~$\mu$g/cm$^2$). The corresponding 2D spectrum (E vs TOF) of the background run is shown in Fig.~\ref{fig_bgr_corr}. As can be seen from the graph, no blob of correlated events is evident in the region around 450~ns.

\subsection{Energy and Angular Acceptance of the System}\label{sub_sec_acceptace}

\begin{figure}
\centering
\includegraphics[width=0.35\textwidth]{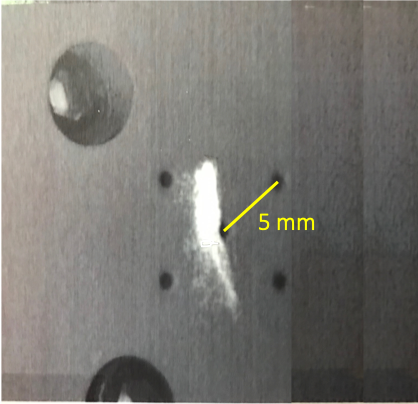}
\caption{Picture of the beam-spot at the position of the HDPE target during the test experiment. The picture was taken by using a beam viewer system that is mounted on the same diagnostics box with the target ladder (see also Fig.~\ref{fig_setup}). The fluorescent screen of the viewer illuminates when the beam ions interact with the material. The five dots on the viewers screen shapes a rectangle that is used for determining the size of the beam-spot. }
\label{fig_beam_spot}
\end{figure}

The maximum energy and angular acceptance of the system was extracted from beam dynamics simulations using the experimental settings from run-1 and run-2 (see Sec.~\ref{sub_sec_dynac_simul}). In these simulations, we considered $^{40}$K ions (Z=19, A=40) emitted by sampling a Gaussian distribution with a mean energy of 124~MeV (the reference beam energy of the test experiment defined by the set magnetic rigidity B$\rho$ of the dipole magnets), and an energy spread with a standard deviation of 5~MeV. Similarly, the angle of these ions was sampled from a Gaussian distribution centered at zero, with an angular spread characterized by a standard deviation of 10~mrad. These initial conditions were chosen such that the emitted ions cover a broad range of angles and energies, surpassing the maximum acceptance of the system that was to be determined. Regarding the size of the beam-spot, we considered a (2$\times$6)~mm elliptical spot that matches the actual size of the beam-spot at the target position (see Fig.~\ref{fig_beam_spot}). Here we assume that within the thickness of the target foil, any changes on the size of the spot are negligible. 

\begin{figure}[t]
\centering
\includegraphics[width=0.45\textwidth]{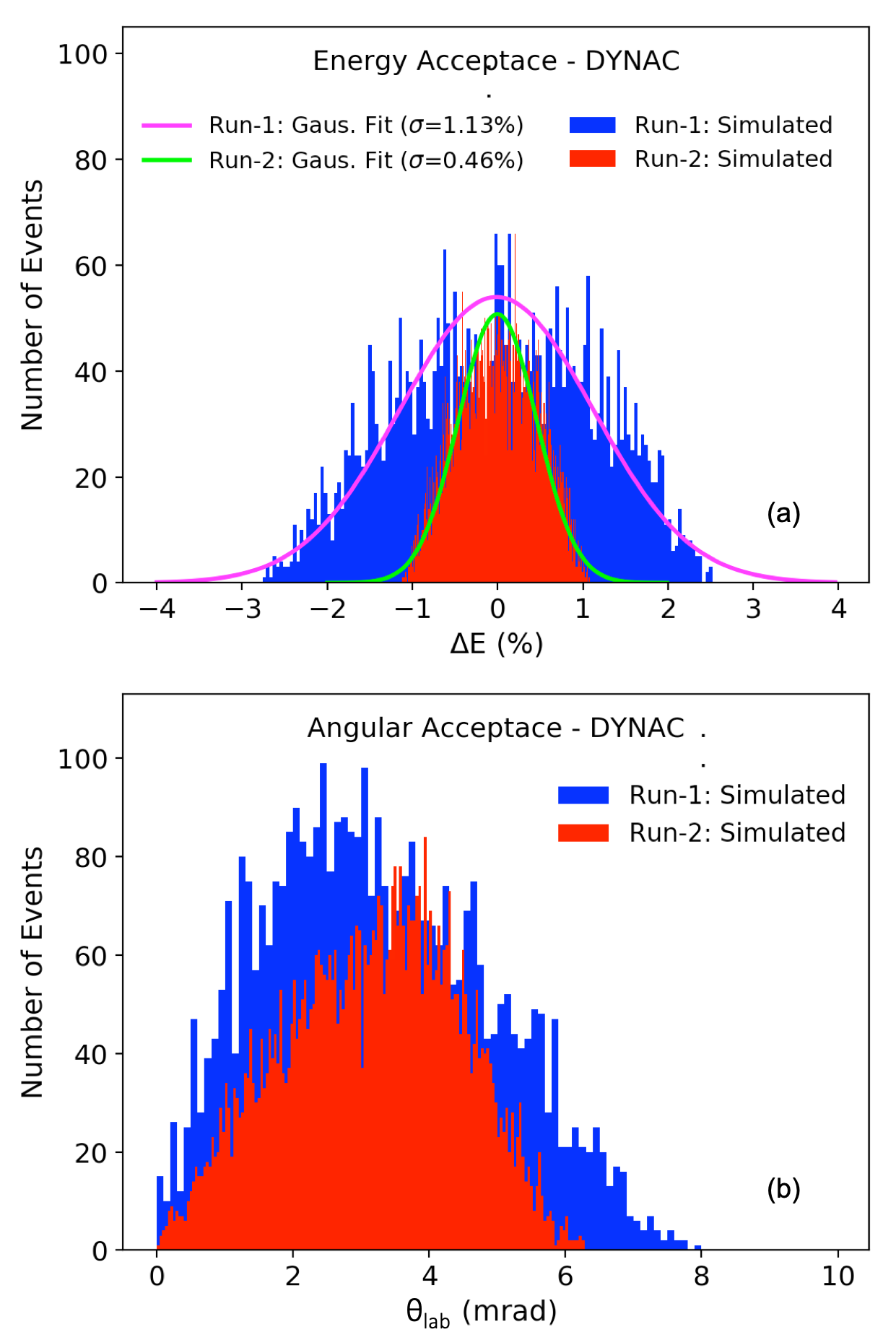}
\caption{Calculated energy and angular acceptance of the system based on the experimental ion-optics settings for run-1 and run-2. The plots display the initial properties of the transmitted $^{40}$K ions through the section of the beamline between the hydrogen target and the DSSD. Details on the initial conditions of the simulation are given in text. (\textbf{a}) Energy distribution of the accepted ions. The distributions show the maximum accepted variation of the ions energy relative to the chosen reference value. (\textbf{b}) Corresponding angular distribution of the accepted ions. }
\label{fig_acceptance}
\end{figure}

In Fig.~\ref{fig_acceptance}, we present the calculated angular and energy distribution of the accepted $^{40}$K ions during run-1 and run-2, based on the DYNAC simulation. In both runs, the maximum angular acceptance was found to be less than 8~mrad, while the energy acceptance relative to the reference energy was limited to about $\pm$2.5\% for run-1, and $\pm$1\% for run-2.

\begin{figure}[t]
\centering
\includegraphics[width=0.48\textwidth]{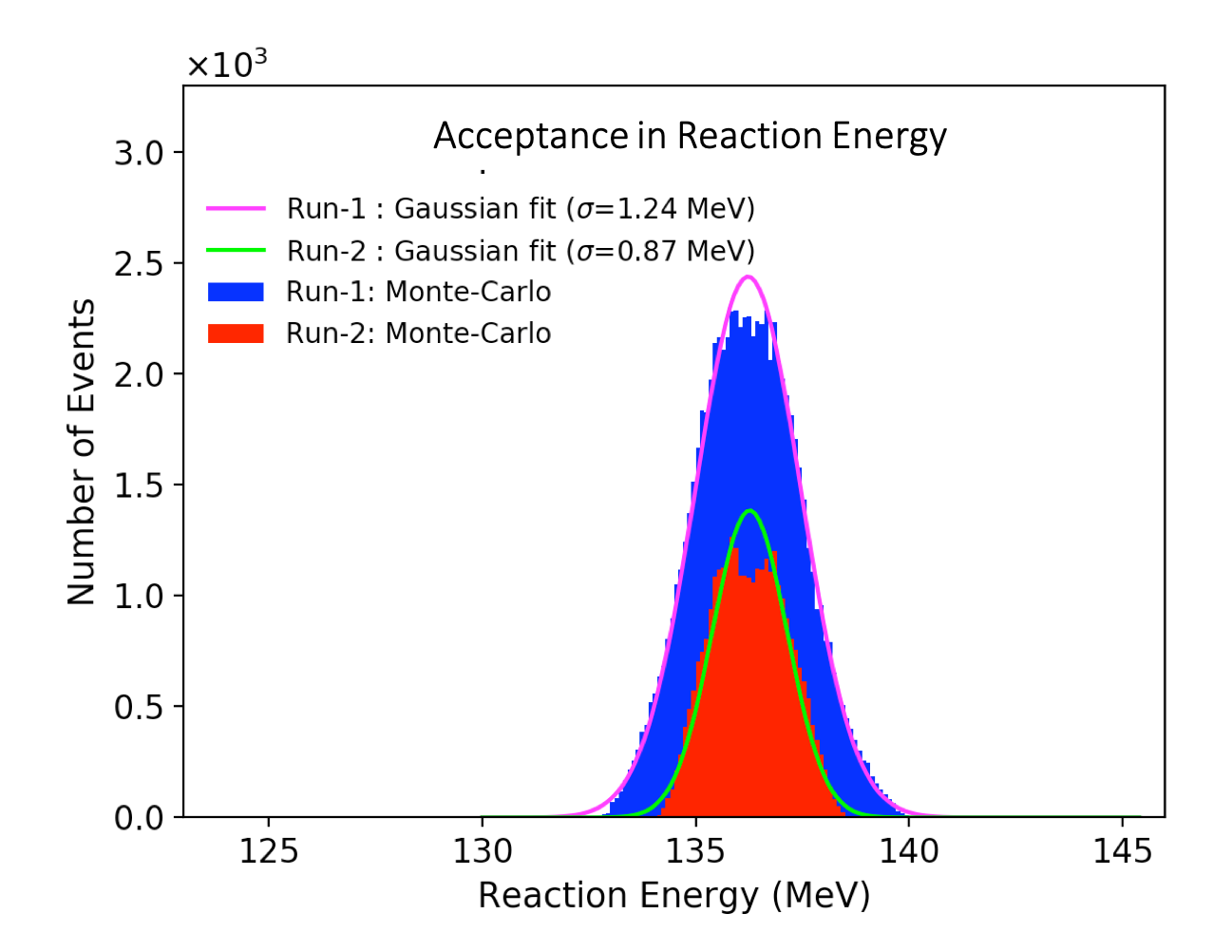}
\caption{Distribution of reaction energies that were "accepted" by the system during the test experiment. The distributions were extracted from Monte-Carlo simulations based on the energy and angular acceptance of the reaction products. The mean energy of these distributions is 136.5~MeV (lab system). The energy uncertainty of the measured cross-sections in each case is given as $\pm$2.35$\sigma$.}
\label{fig_reac_acc}
\end{figure}

The limited acceptance of the system affects both the yields of the measured reaction products, and the energy width of the measured cross-sections, i.e., the energy range over which the measured cross-section is averaged (energy bin). This range typically depends on the energy spread of the incoming beam and the energy loss of the beam inside the target. Due to energy dispersion in beamline optics, when the acceptance of the system is limited, the width of the energy bin of a data point is not simply given by the energy loss or the energy spread of the incoming beam, but must be calculated based on the system's acceptance. By using the calculated energy and angular spread of the reaction products (details on these Monte-Carlo simulations are given in Sec.~\ref{sub_sec_dynac_simul}), as well as the acceptance of the system for run-1 and run-2, we extracted the distribution of reaction energies that were ``accepted" during the test experiment. The results of this analysis are shown in Fig.~\ref{fig_reac_acc}, where each distribution is fitted with a Gaussian function. Based on this approach, the final energy uncertainty of the data points in this experiment, given as the half-width of the energy bin (2.35$\sigma$), corresponds to $\Delta$E values of $\pm$2.13\% in run-1, and $\pm$1.5\% in run-2.

\subsection{Transmission of $\,^{40}$K Ions to the DSSD}\label{sub_sec_trans_prod}

Having a detailed picture of the reaction kinematics and taking into account the acceptance of the system during the test experiment (see Fig.\ref{fig_acceptance}), it is evident that the vast majority of the nuclei from (p,n$_0$) and (p,n$_1$) channels fall outside our acceptance window (see Fig.~\ref{fig_kinematics}). In both runs (run-1 and run-2), the angular acceptance of the system was less than 8~mrad, while the acceptance in energy was no more than $\pm$3 MeV from the mean value. Combining this information with the results from kinematics, we estimated that the portion of the products from (p,n$_0$) and (p,n$_1$) that fits in the acceptance window is less than 0.1\%. As a result, the contribution from these two channels to the measured cross-sections is negligible. Furthermore, according to the experimental data on the partial cross-sections for the $^{40}$Ar(p,n)$^{40}$K reaction~\cite{Gastis2020a}, the contribution of the (p,n$_3$) channel is also negligible compared to the dominant (p,n$_2$) channel. Therefore, during the test experiment only ions from the (p,n$_2$) channel were reaching the DSSD. If this was not the case and ions corresponding to more channels were reaching the DSSD, the situation would be significantly more complicated and the accuracy of the results would depend on our ability to disentangle the yields of the individual channels.

\begin{figure}
\includegraphics[width=0.48\textwidth]{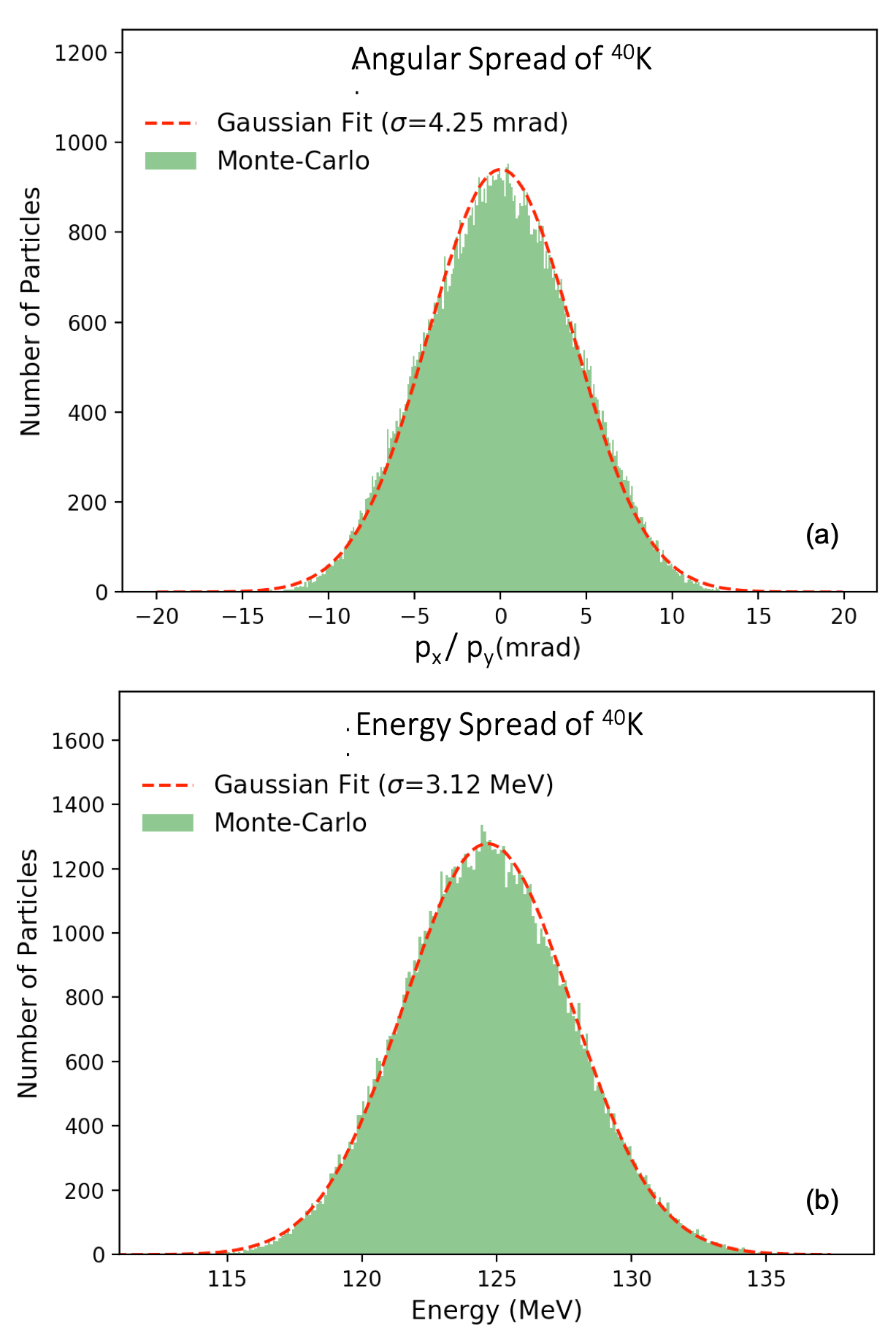}
\caption{Energy and angular characteristics of the reaction products according to our Monte-Carlo calculations. (\textbf{a}) Angular spread of $^{40}$K relative to the horizontal (x) and vertical (y) axis. The azimuthal angle of the ions is analyzed into the x and y components of Eq.~\ref{eq_theta}. Due to the symmetry of the system, the p$_x$ and p$_y$ distributions are almost identical. For this reason we present them as a single distribution. (\textbf{b}) Energy spread of $^{40}$K at the exit of the target. The energy loss of the beam ions in the target foil, resulted in an adequately averaged energy distribution for the reaction products. A spread in the emittance parameters of one  standard deviation based on the above distributions was used to extract the emittance ellipsoids for the products, and subsequently, to calculate their transmission through the beamline. }
\label{fig_prod_em}
\end{figure}

The transmission of $^{40}$K through the beamline was extracted from the beam dynamics simulations. To model the products' emittance in DYNAC, we provided input parameters according to the angular and energy distributions of the $^{40}$K ions from the (p,n$_2$) channel. Figure~\ref{fig_prod_em} shows the angular and energy distributions of the ions based on the Monte-Carlo analysis method that we discussed in Sec.~\ref{sub_sec_dynac_simul}. For the beam-spot size, we used the same conditions as in the calculation of the system's acceptance. Based on the above data, we were able to extract realistic input parameters for generating the initial phase-space of the products beam. 

The results of the beam dynamics simulations are summarized in Table~\ref{tab:transm}. In both runs, the transmission of products was much less than 10\%, indicating that the applied ion optics during the test experiment were not properly optimized for product transmission. Nevertheless, by implementing the extracted transmissions in Eq.~\ref{eq_cross_sec}, we were able to obtain the reaction cross-section for each run and compare the final results with the literature data. 

The estimated errors on the calculated transmissions with this method are on the order of 12\%. This error was calculated based on the expected uncertainties on the initial energy and angular spread of the products. The theoretical models that we used for extracting these quantities are accurate within 5\%~\cite{Geissel:2002,Ziegler:125402}. By modifying the initial emittance of the products according to this uncertainty, the maximum variations of the calculated transmissions with DYNAC were extracted. This method was also used to estimate the corresponding errors for the unreacted beam that we discuss in the following section.

\begin{table}
\centering
\caption{Transmission of reaction products and unreacted beam through the beamline. The calculated values were extracted from beam dynamics simulations with DYNAC. The last row includes data for the transmission of the incoming beam without passing through the target foil. To cross-check the accuracy of the calculated transmissions for the unreacted beam, the obtained values are tested against experimental data that we collected using Faraday cups. }
\label{tab:transm}
\begin{tabular*}{0.95\columnwidth}{cccc}

 \rule{0pt}{0.5cm}
 &  \multicolumn{2}{c}{Calculated}  &  \multicolumn{1}{c}{Experimental}  \\
\cline{2-3} 

 \rule{0pt}{0.4cm} Data Set  & $^{40}$K (\%) & $^{40}$Ar (\%) & $^{40}$Ar (\%) \\ \hline
 \rule{0pt}{0.36cm}
 
 Run-1 & 1.35 $\pm$ 0.16 & 4.8 $\pm$ 0.5  & 5.0 $\pm$ 1.0 \\
 Run-2 & 5.90 $\pm$ 0.71 & 18.8 $\pm$ 1.5 & 21.2 $\pm$ 4.3 \\
 w/o foil & N/A & 69.5 $\pm$ 4.0 & 66.0 $\pm$ 6.6 \\

\end{tabular*}
\end{table}

\subsection{Experimental Validation of the Calculated Transmissions}\label{sub_sec_transmission_beam}

To test the validity of the above method in calculating the transmission of nuclei through the beamline, we checked whether this model can reproduce the transmission of the unreacted $^{40}$Ar beam. During the test experiment, the transmission of the unreacted beam was determined using the beam diagnostics devices of ReA3. By using these data as a reference, we repeated the above analysis for the case of the $^{40}$Ar beam. In the first part of this analysis, we tested the accuracy of the simulated transmission of the $^{40}$Ar$^{+14}$ beam without passing through the HDPE foil. For the incident angles p$_x$ and p$_y$ at the target position, we assumed Gaussian distributions with a mean value of 0~mrad and a FWHM of (2.0$\pm$0.25)~mrad. This angular distribution was estimated during the experiment while tuning the incoming $^{40}$Ar$^{14+}$ beam. Regarding the energy spread of the beam, we assumed a mean energy of 140.63 MeV (3.52 MeV/nucleon) and a FWHM of (0.87$\pm$0.04) MeV. The FWHM of the distribution was determined experimentally using the dipole magnet upstream the 4-jaw system (see Fig.\ref{fig_setup}) and a Faraday cup. The exact method of measuring the energy spread is described in the next paragraph. By implementing the above beam characteristics in DYNAC we calculated the transmission of the incoming beam through the section between the target position and the DSSD. In this simulation, we implemented the experimental ion-optics settings that were used while measuring the transmission of the $^{40}$Ar$^{+14}$ beam along the beam line. The simulated transmission was found to be (69.5$\pm$4.0)\% which is in a good agreement with the experimental value of (66$\pm$6.6)\%. The error in the calculated transmission was extracted by using the method we described in Sec.~\ref{sub_sec_trans_prod}, taking into account the estimated uncertainties on the initial angular and energy spread of the beam ions. 

\begin{figure}
\includegraphics[width=0.45\textwidth]{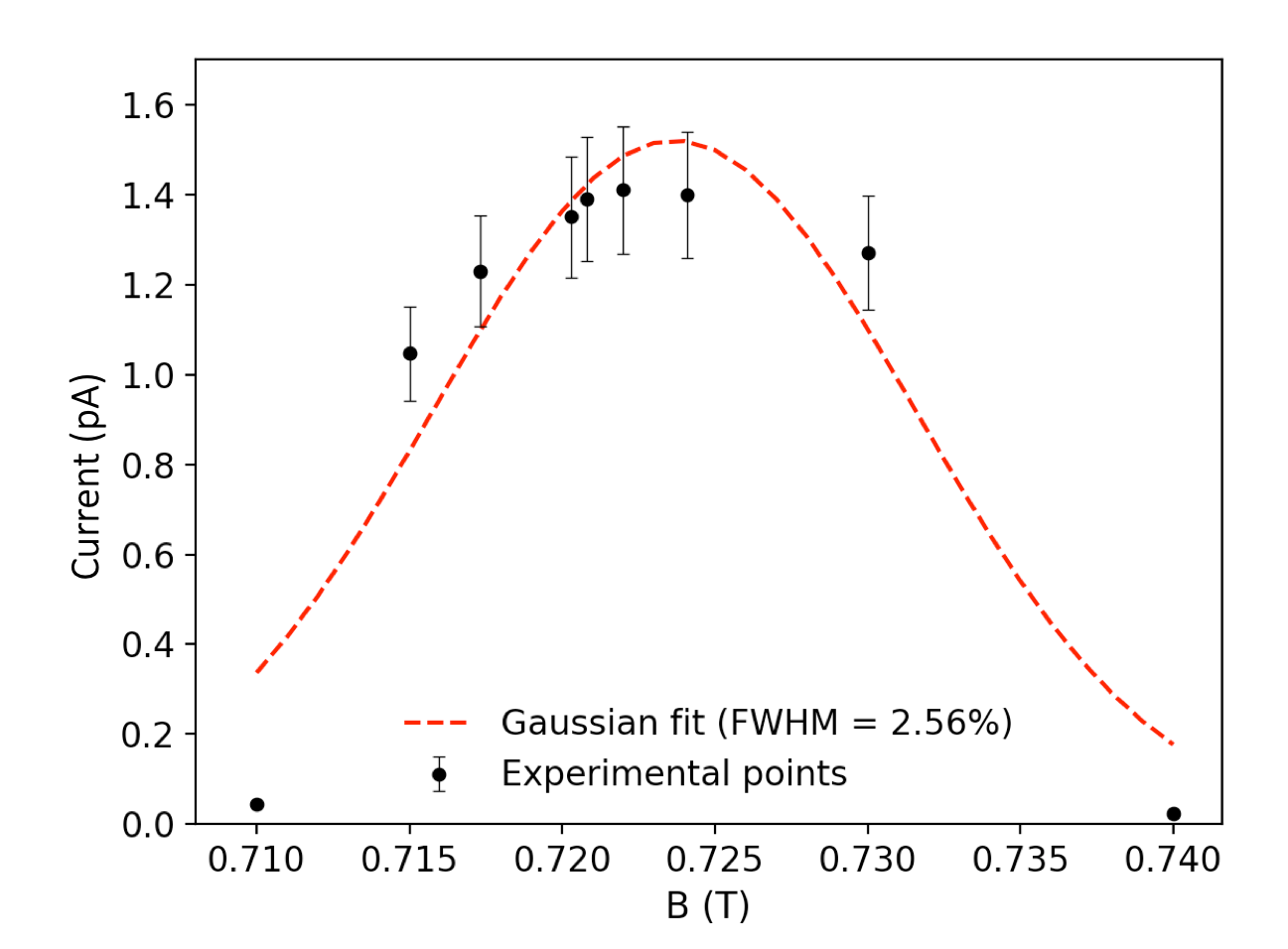}
\caption{Momentum spread of the unreacted $^{40}$Ar beam, determined by using the dipole magnet upstream the 4-jaw slits system. The plot shows the measured beam current behind the slits as a function of the applied magnetic field. During the measurement, the slits opening was 0.5~mm. The energy spread of the beam was obtained by fitting the above experimental data with a Gaussian function and converting the spread in magnetic field to energy spread using Eq.~\ref{eq_spread_E}. }
\label{fig_spread}
\end{figure}

In the second part of the analysis, we calculated the transmission of the beam after passing through the HDPE foil. For this part, we used the experimental ion-optics settings of run-1 and run-2. To obtain the angular spread of the beam after passing through the HDPE foil, we performed the same Monte-Carlo analysis as described in the previous paragraphs. To describe the angular straggling of $^{40}$Ar in HDPE we used the SW-theory, while for the initial angular spread we assumed a FWHM of (2.0$\pm$0.2)~mrad (as before). The energy spread of the unreacted beam after passing the HDPE target was determined experimentally using a dipole magnet and a Faraday cup. Specifically, we measured the intensity of the beam after the 4-jaw system (see Fig.~\ref{fig_setup}), as a function of the field of the dipole magnet that is located upstream of the slits. During this measurement, the opening width of the slits was about 0.5~mm, while the Faraday cup was mounted right behind the slits. In this way, we were able to measure the intensity profile of the beam in terms of the B field. The results of this measurement are presented in Fig.~\ref{fig_spread}. To obtain the corresponding energy spread, we used the relationship that connects the amplitude of the magnetic field and the kinetic energy of the ions. The kinetic energy $E$ of an ion with mass $m$ and charge $q$ that is moving inside a magnetic field $B$ ($B$ is perpendicular to the ion's velocity), is equal to:
\begin{equation}
 E = \frac{(B\,\rho\, q)^2}{2m}
\label{eq_energy}
\end{equation}
where $\rho$ is the radius of the circular trajectory of the moving ion. By differentiating the above equation, we get:
\begin{equation}
 \frac{\delta E}{E} = 2 \, \frac{\delta B}{B}
\label{eq_spread_E}
\end{equation}
By using Eq.~\ref{eq_spread_E}, we extracted FWHM of (5.7$\pm$0.1)~MeV for the energy distribution of the unreacted $^{40}$Ar at the exit of the target. The mean energy of the ions was set to 131.67~MeV, taking into account the energy loss in the HDPE. By implementing the above beam characteristics in DYNAC, we calculated a transmission of (4.8$\pm$0.4)\% for run-1, and (18.8$\pm$1.5)\% for run-2. The corresponding experimental values are (5.0$\pm$1.0)\% and (21.2$\pm$4.3)\%, respectively. By comparing the two results we observe an excellent agreement between them, as in both cases, the calculated values are within the experimental errors. These results are tabulated in Table~\ref{tab:transm}. 

The errors in the calculated transmissions were extracted based on the uncertainties (experimental and theoretical) on the initial angular and energy spread of the beam ions. For the experimental values we adopted an uncertainty of 20\%. The experimental transmissions were determined by measuring the beam intensity at the DSSD position with a femto-ampere meter, and at the target position with a Faraday cup. The measurements were repeated using various charge-states of the unreacted beam. The error of these measurements was estimated after taking into account the variations that we observed in the measurements with different charge-states, as well as the uncertainty due to the measurement of very low intensities.

\subsection{Experimental Validation of the Calculated Charge-State Distributions}\label{sub_sec_charge_states}

We performed measurements on the charge-state distribution of the unreacted $^{40}$Ar$^{14+}$, $^{40}$Ar$^{15+}$, and $^{40}$Ar$^{16+}$ charge-states using a Faraday cup. These data were used as a reference in order to check the accuracy of the semi-empirical models that we used for the calculation of the charge-state distribution of $^{40}$K. The results of these measurements are presented in Table~\ref{tab:results_charge_states}, along with the theoretical calculations that we discussed in Sec.~\ref{sub_sec_charge_fractions}. Due to the low precision of the measurements with the Faraday cups (affected by the low transmission of the ions) an uncertainty of 15\% was adopted for the experimental charge-state fractions. For the fractions of the most intense charge-states, the agreement between the experimental data and the calculations is very good. The enhanced discrepancy on the 14+ charge-state is attributed to shell effects. This charge-state coincides with the s-p orbital transition in Ar, and therefore, its intensity is much lower compared to the intensity calculated from our model (see Ref.~\cite{Gastis:2015cfa}). 

\begin{table}
\centering
\caption{Charge-state distributions of $^{40}$Ar and $^{40}$K in HDPE. The theoretical charge-state fractions were extracted by using the Schiewitz model~\cite{Schiwietz:2004}. To cross-check the accuracy of the calculated fractions, the theoretical values are tested against experimental data for the unreacted beam that we collected with a Faraday cup.}
\label{tab:results_charge_states}
\begin{tabular*}{0.95\columnwidth}{cccc}

 \rule{0pt}{0.5cm}
 &  \multicolumn{2}{c}{Calculated}  &  \multicolumn{1}{c}{Experimental}  \\
\cline{2-3} 
 \rule{0pt}{0.4cm} Charge  & $^{40}$K~(\%) & $^{40}$Ar~(\%) & $^{40}$Ar~(\%) \\ \hline
 \rule{0pt}{0.36cm}
 
 14+ &  9.5 $\pm$ 1.0 & 20.0 $\pm$ 2.0  & 16.0 $\pm$ 2.4  \\
 15+ & 28.0 $\pm$ 2.8 & 37.0 $\pm$ 3.7  & 35.0 $\pm$ 5.3  \\
 16+ & 35.5 $\pm$ 3.6  & 29.0 $\pm$ 2.9  & 27.0 $\pm$ 4.1  \\

\end{tabular*}
\end{table}

\subsection{Neutron Angular Distribution}\label{sub_sec_solid_ang}

\begin{figure}
\includegraphics[width=0.98\linewidth]{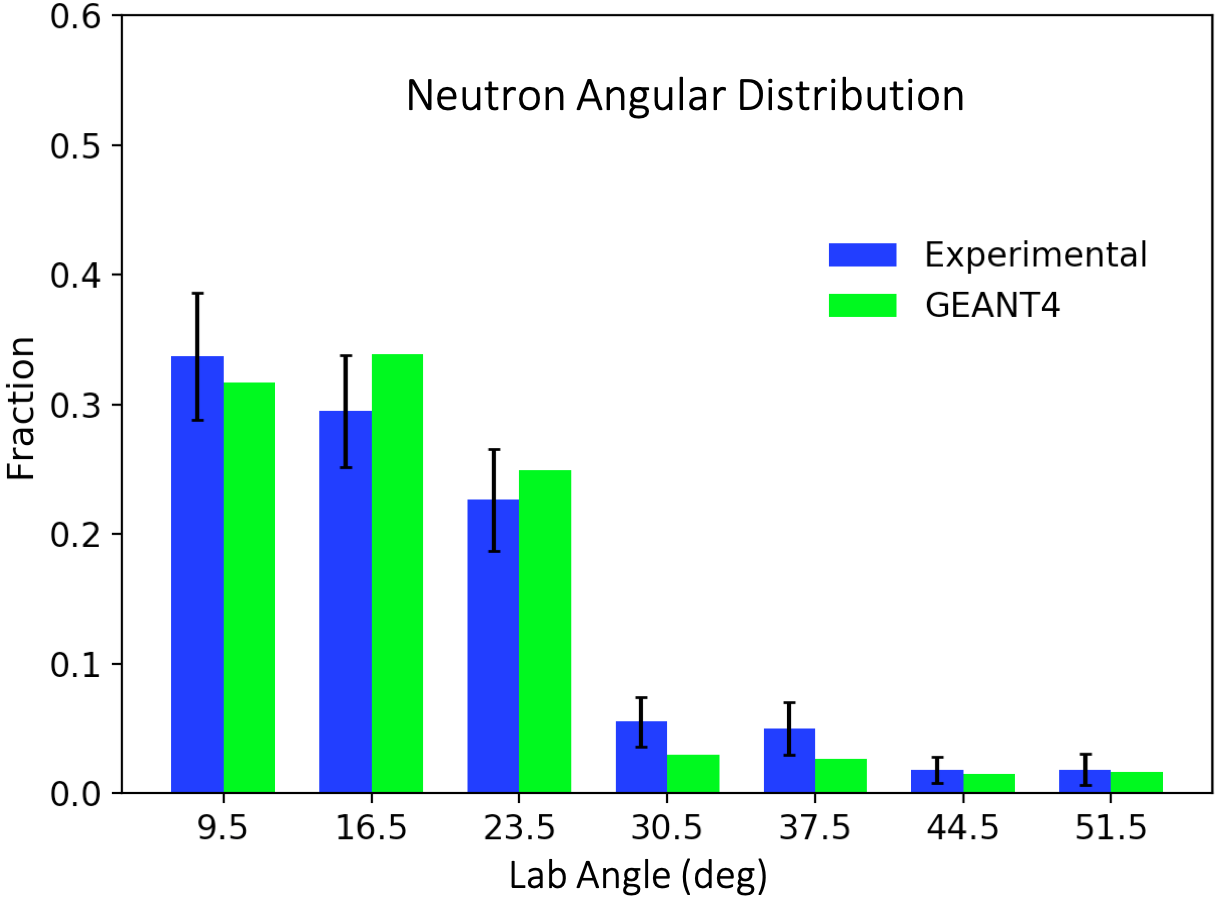}
\caption{Neutron angular distribution of the $^{40}$Ar(p,n$_2$) reaction from this work. The angles are given relative to the beam axis and correspond to the center of each bar. The vertical axis shows the portion of the detected neutrons that ended up at each angle. (\textit{blue bars}) Experimental angular distribution based on the cumulative data from run-1 and run-2. (\textit{green bars}) Simulated angular distribution using GEANT4. The simulation takes into account the reaction kinematics, the intrinsic efficiency of each bar, and the scattering of neutrons on the various beamline components and the LENDA bars. Details on this GEANT simulation are discussed in Sec.~\ref{sub_sec_solid_ang_analysis}.}
\label{fig_neut_dist}
\end{figure}

 Figure~\ref{fig_neut_dist} shows the experimentally determined neutron angular distributions for the $^{40}$Ar(p,n$_2$) channel. The horizontal axis corresponds to the mean angle covered by each LENDA bar in the laboratory frame. The bars were arranged symmetrically around the beam axis (see Fig.~\ref{fig_lenda}), therefore, each angle in this plot includes data from two bars. The corresponding fractions on the vertical axis are calculated by dividing the number of detected neutrons in each angle with the total number of detected neutrons. The presented angular distribution combines cumulative data from both runs.

As expected from the reaction kinematics and the geometry of the system, the vast majority of neutrons were detected at angles below 25$^\circ$. The experimental data show that about 14\% of the detected neutrons are distributed at angles larger than 25$^\circ$, even though no neutrons from the (p,n$_2$) channel are expected in that region according to kinematics. These counts are associated with scattered neutrons on the stainless steel components of the beamline, as well as on the various LENDA bars (cross-talk). That is evident from the GEANT4 simulations discussed in Sec.~\ref{sub_sec_solid_ang_analysis} that were performed for extracting the solid angle of LENDA. The corresponding angular distribution from those simulations is included in Fig.~\ref{fig_neut_dist}. The good agreement between the experimental data and the GEANT simulation indicates that the calculated solid angle coverage and the intrinsic efficiency of LENDA are adequately accurate, and any underlying systematic uncertainties associated with those quantities are well withing the estimated errors.

\subsection{Time-Spread of Correlated Events and DSSD Timing Resolution}\label{sub_sec_time_res}

\begin{figure}
\includegraphics[width=0.48\textwidth]{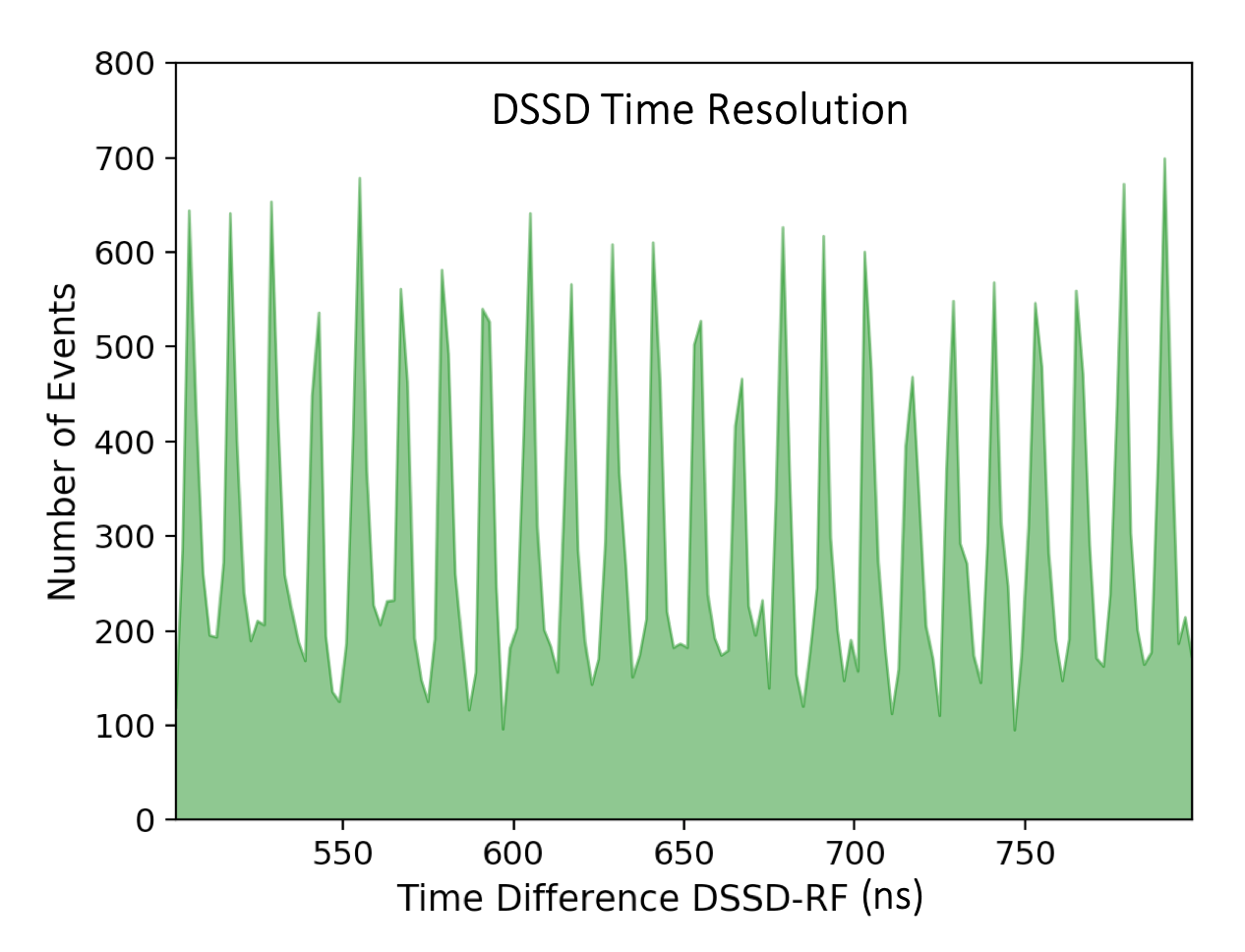}
\caption{Time resolution of the DSSD, as measured with the $^{40}$Ar beam. This histogram shows the time difference between the signal of the DSSD and a reference RF signal from the re-accelerator. The RF signal has a period of 12~ns resulting in the formation of consecutive peaks in the spectrum that were observed with a good resolution. From this plot, we extract an upper limit of 3.5~ns for the timing resolution of the DSSD. }
\label{fig_rf_dssd}
\end{figure}

The timing resolution of the system was dominated by the DSSD's timing response. To get an estimate of this resolution, we analyzed a TOF spectrum of the time difference between the $^{40}$Ar beam, and the reference RF signal from the re-accelerator buncher. The spectrum is presented in Fig.~\ref{fig_rf_dssd}. The well-separated peaks reflect the 12~ns period of the RF signal. Considering the average width of these peaks, an upper limit of 3.5~ns was extracted for the time resolution of the DSSD. To determine the detector's resolution more precisely, the time coordinates of the beam ions must be taken into account. The task of measuring the resolution with higher precision was not within the goals of this study, hence we limit ourselves in this work to quoting only the upper limit.

\begin{figure}
\includegraphics[width=0.48\textwidth]{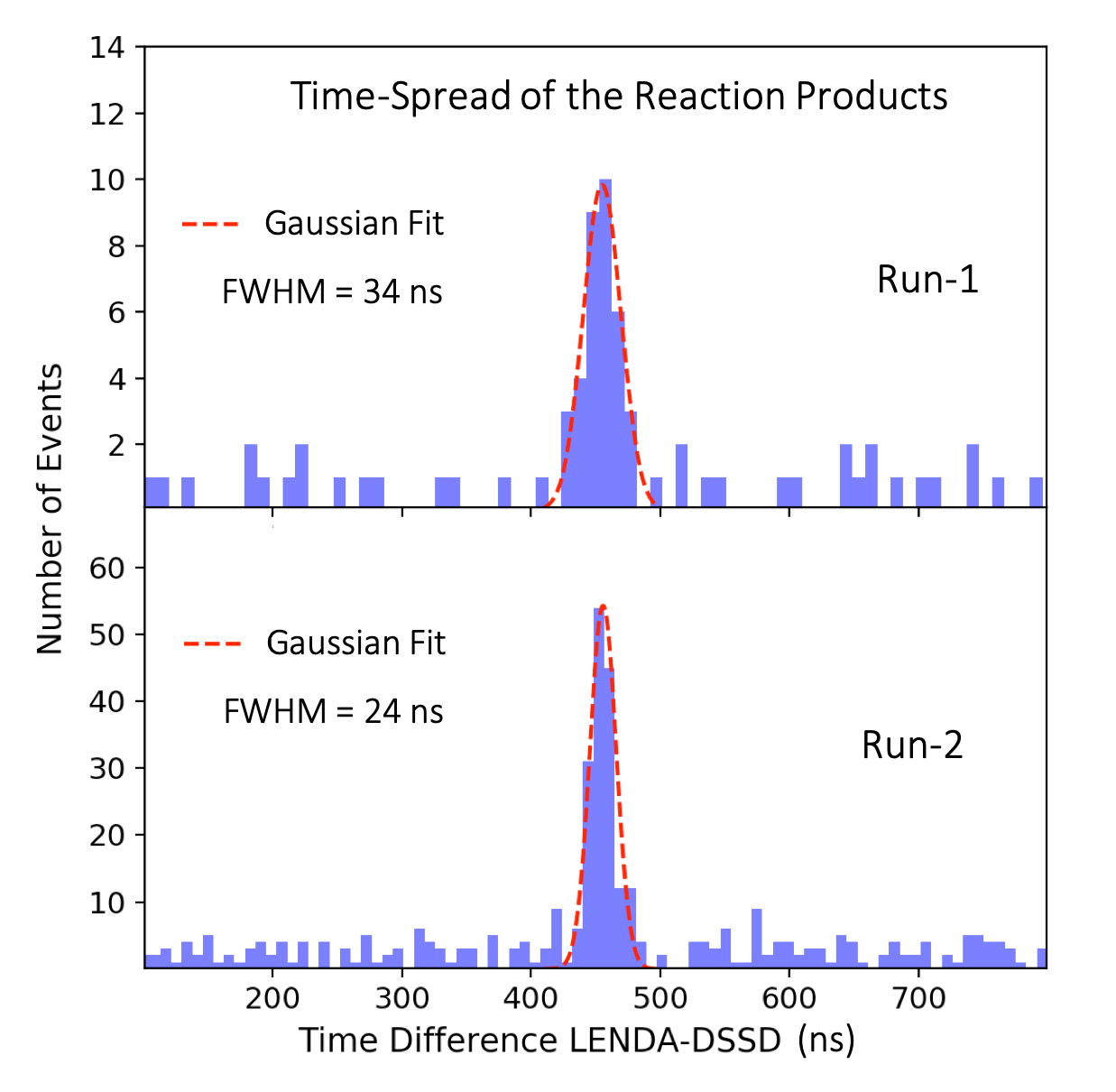}
\caption{Time-of-flight peaks of correlated events. The plots were extracted from the 2D spectra presented in Figs.~\ref{fig_run1_corr} and ~\ref{fig_run2_corr}, after subtracting a linear background. To obtain their FWHM, the peaks were fitted with a Gaussian function. The extracted widths were found to be consistent with the energy spread of neutrons on LENDA, and the energy acceptance of the system for the recoils.}
\label{fig_time_width}
\end{figure}

While the time resolution of our detectors is relatively high, the average width of the TOF peaks in the spectra is quite large. For example, the TOF peaks in Figs.~\ref{fig_run1_corr} and \ref{fig_run2_corr} have a FWHM of 34~ns (run-1), and 24~ns (run-2), respectively. These FWHM were extracted by fitting the experimental data with a Gaussian function, as shown in Fig.~\ref{fig_time_width}. The enlargement of the FWHM on the TOF spectra is due to the time-spread of the events (time difference between correlated events in the two detectors) and depends on the energy spread of the detected particles.
According to the Monte-Carlo reaction kinematics (see Sec.~\ref{sub_sec_solid_ang_analysis}), the neutron energies in these measurements spanned between 1.5 and 6~MeV. Considering the geometry of LENDA and its distance from the target, the corresponding TOF spread of the detected neutrons is $\sim$14~ns. Regarding the TOF spread of the $^{40}$K ions, this can be estimated from the calculated energy acceptance of the system. The kinetic energy of a particle with mass $m$, that covers a distance $L$ in time $\tau$, is given from the equation:
\begin{equation}
E = \frac{1}{2} \, m \, \left(\frac{L}{\tau} \right)^2
\label{eq_kin_energy}
\end{equation}
By differentiating the above equation, we find that:
\begin{equation}
 \frac{\delta E}{E} = 2 \, \frac{\delta \tau}{\tau}
\label{eq_spread_in_time}
\end{equation}
Using Eq.~\ref{eq_spread_in_time} and considering an energy spread of $\pm$2.5\% for run-1, and $\pm$1.0\% for run-2 (see Fig.~\ref{fig_acceptance}) we obtain a time spread of 18.1~ns for the $^{40}$K ions during run-1, and 7.2~ns during run-2. The mean TOF in these calculations is 725~ns (the flight path is 17.8 m and the mean energy of $^{40}$K is 124~MeV). Adding together the estimated time-spread of neutrons and $^{40}$K, and the time resolution of LENDA and the DSSD, the expected time-spread of the reaction products is 36.1~ns for run-1 and 25.2~ns for run-2. Within an uncertainty of 6\%, these numbers are in agreement with the FWHM of the experimental TOF peaks (see Fig.~\ref{fig_time_width}), indicating that the calculated energy acceptance of the system with the beam dynamics simulation is adequately accurate.

\subsection{Reaction cross-section and Comparison with Literature}\label{sub_sec_results}

By using the experimental input parameters, we extracted the partial cross-section of the $^{40}$Ar(p,n$_2$) channel for run-1 and run-2. The results are given in Table~\ref{tab:results_sigma}. The corresponding errors on the total cross-sections are extracted from error propagation in Eq.~\ref{eq_cross_sec}.

\begin{figure}
\includegraphics[width=0.48\textwidth]{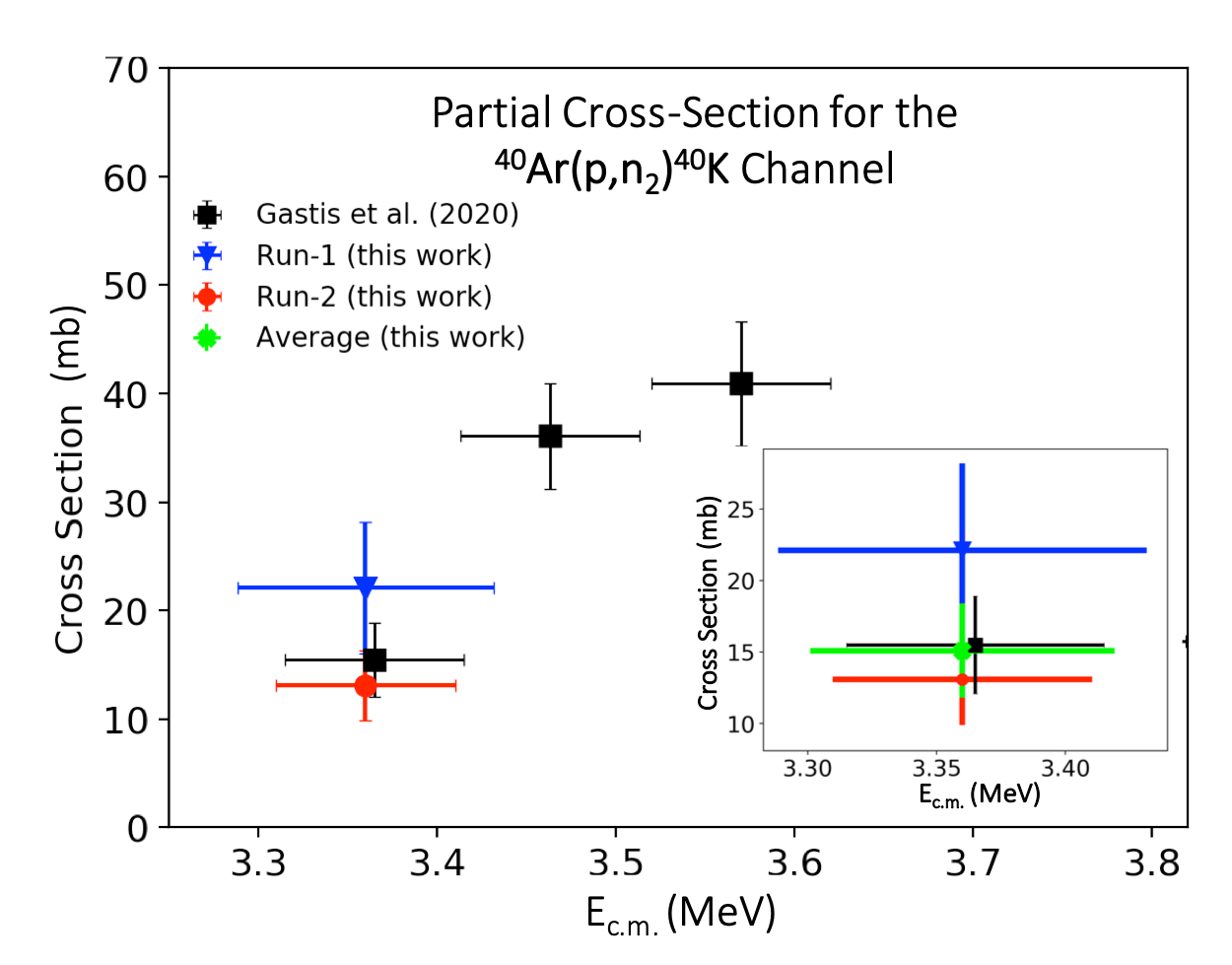}
\caption{Comparison of the measured cross-sections in run-1 (blue triangles) and run-2 (red circles), with literature data (black squares). Reference data on the partial cross-sections were retrieved from~\cite{Gastis2020a}. Within the experimental errors, the extracted cross-sections are in a good agreement with the reference data. At 3.36~MeV, the data point from run-1 is above the other two points due to the larger energy uncertainty of that point (larger energy bin size) and the increasing trend of the cross-section evident in the literature data from \citet{Gastis2020a}. An excellent agreement with the literature data is also observed for the weighted-average cross-section of the two runs, which is shown in the zoom-in graph (right-corner).}
\label{fig_result_cs}
\end{figure}

The results are plotted and compared to literature data from Ref.~\cite{Gastis2020a} in Fig.~\ref{fig_result_cs}. The cross-sections extracted from both runs agree within the limits of their corresponding uncertainties with the reference data. The difference in the measured values between the two runs are attributed to the different energy uncertainties of each measurement (for details see Sec.~\ref{sub_sec_acceptace}). The average cross-section over the energy range covered by the measurement in run-1 was significantly higher than the one in run-2 that covered a smaller energy range. This result is in agreement with the measured cross-section for the $^{40}$Ar(p,n$_2$) channel by~\cite{Gastis2020a}. Two broad resonance structures corresponding to isobaric analog states of $^{41}$Ar exist at 3.3 and 3.5 MeV are boosting the average cross-section for the measurement in run-1 to a higher value. Part of that strength is shown with black squares in Fig. \ref{fig_result_cs} for comparison. 
In the same graph, we present the weighted-average of the two runs, which was found to be in excellent agreement with the reference data at $\sim$3.36~MeV. The energy uncertainty for the average cross-section was calculated as:
\begin{equation}
\overline{\Delta E} = \sqrt{ w_1 \Delta E_1^2 + w_2 \Delta E_2^2 }
\label{eq_ave_cs}
\end{equation}
where $\Delta E_1$, $\Delta E_2$ are the energy uncertainties (energy bins) in run-1 and run-2, and w$_1$,  w$_2$ are weighting  factors extracted from the statistics of each run. Specifically, the weighting  factor w$_1$ was calculated as: w$_1$= Y$_1$/(Y$_1$+Y$_2$), where Y$_1$, Y$_2$ are the measured neutron yields during run-1 and run-2, respectively, while w$_2$= 1--w$_1$.

\begin{table}[h]
\centering
\caption{Total cross-section of the $^{40}$Ar(p,n$_2$)$^{40}$K reaction channel. In this table, we provide the measured cross-sections from run-1 and run-2, as well as the weighted-average of the two runs. Detailed discussion on these results are given in the text.  }
\label{tab:results_sigma}
\begin{tabular*}{0.8\columnwidth}{ccc}
 \rule{0pt}{0.5cm}
 Data-Set  & E$_{c.m.}$ (MeV) & $\sigma$ (mb) \\ \hline
 \rule{0pt}{0.36cm}
 Run-1 & 3.36 $\pm$ 0.08 & 21.9 $\pm$ 6.3 \\ 
 Run-2 & 3.36 $\pm$ 0.05 & 13.0 $\pm$ 3.3  \\ 
 Average & 3.36 $\pm$ 0.06 & 14.9 $\pm$ 2.9  \\ 
 Gastis et al.~\cite{Gastis2020a} & 3.365 $\pm$ 0.051 & 15.0 $\pm$ 3.2  \\
\end{tabular*}
\end{table}

\subsection{Uncertainty Estimates}
The cross-sections from the proof-of-principle experiment are given with an uncertainty of 29\% for run-1, and 26\% for run-2. These errors were calculated by adding all the uncertainties of the associated quantities in Eq.~\ref{eq_cross_sec} in quadrature. These uncertainties are summarized in Table~\ref{tab:errors}. The final errors on the cross-sections were dominated by the errors of the target areal density ($\sim$10\%), the beam current integration ($\sim$10\%), the charge-state fractions ($\sim$10\%), the calculated transmissions ($\sim$12\%), and the intrinsic neutron efficiency of LENDA ($\sim$11\%). In addition to the above sources of error, the uncertainty in run-1 was also affected by the low statistics in that measurement. While the final errors are relatively large, future improvements on the experimental setup can significantly reduce them, enabling measurements within an uncertainty of 20\%.

One of the dominant uncertainty factors, the transmission of ions along the beamline, can be determined with a much better precision for values larger than 10\%, i.e. if the ion-beam optics are properly optimized; the current transmission was always less than $\sim$6\% in either of the two runs. The second dominant factor has to do with the precision of our knowledge of the charge-state fraction. This contribution to the error can be significantly reduced by a detailed measurement of the charge-state distribution during an experiment. In the test runs, the low transmission of the beamline prevented as from taking high-precision data on the charge-state distributions, but this will not be an issue in future work.

\begin{table}
\centering
\caption{Estimated uncertainties of the various quantities in Eq.~\ref{eq_cross_sec}, that were used for the extraction of the final cross-sections.
}
\label{tab:errors}
\begin{tabular*}{0.8\columnwidth}{ccc}

 \rule{0pt}{0.5cm}
 &  \multicolumn{2}{c}{Uncertainty (\%)}  \\
\cline{2-3} 
 \rule{0pt}{0.4cm} Quantity  & Run-1 & Run-2  \\ \hline
 \rule{0pt}{0.36cm}
 
Products' yield (Y$_n$) &  15.2 & 8.2    \\
Beam current integral (I$_b$) & 10.0 & 10.0    \\
Target's areal density (N$_t$) & 10.0 & 10.0    \\
LENDA solid angle ($\omega_n$) & 5.5 & 5.5    \\
Total neutron efficiency ($\epsilon_n$) & 11.0 & 11.0  \\
Charge-state fraction (F$_q$) &  10.0 & 10.0 \\
Products' transmission (T$_p$)  &  12.0 & 12.0 \\ \hline
Cross-Section ($\sigma$) &  28.8 & 25.7 \\
\end{tabular*}
\end{table}

\begin{figure}[h]
\includegraphics[width=0.48\textwidth]{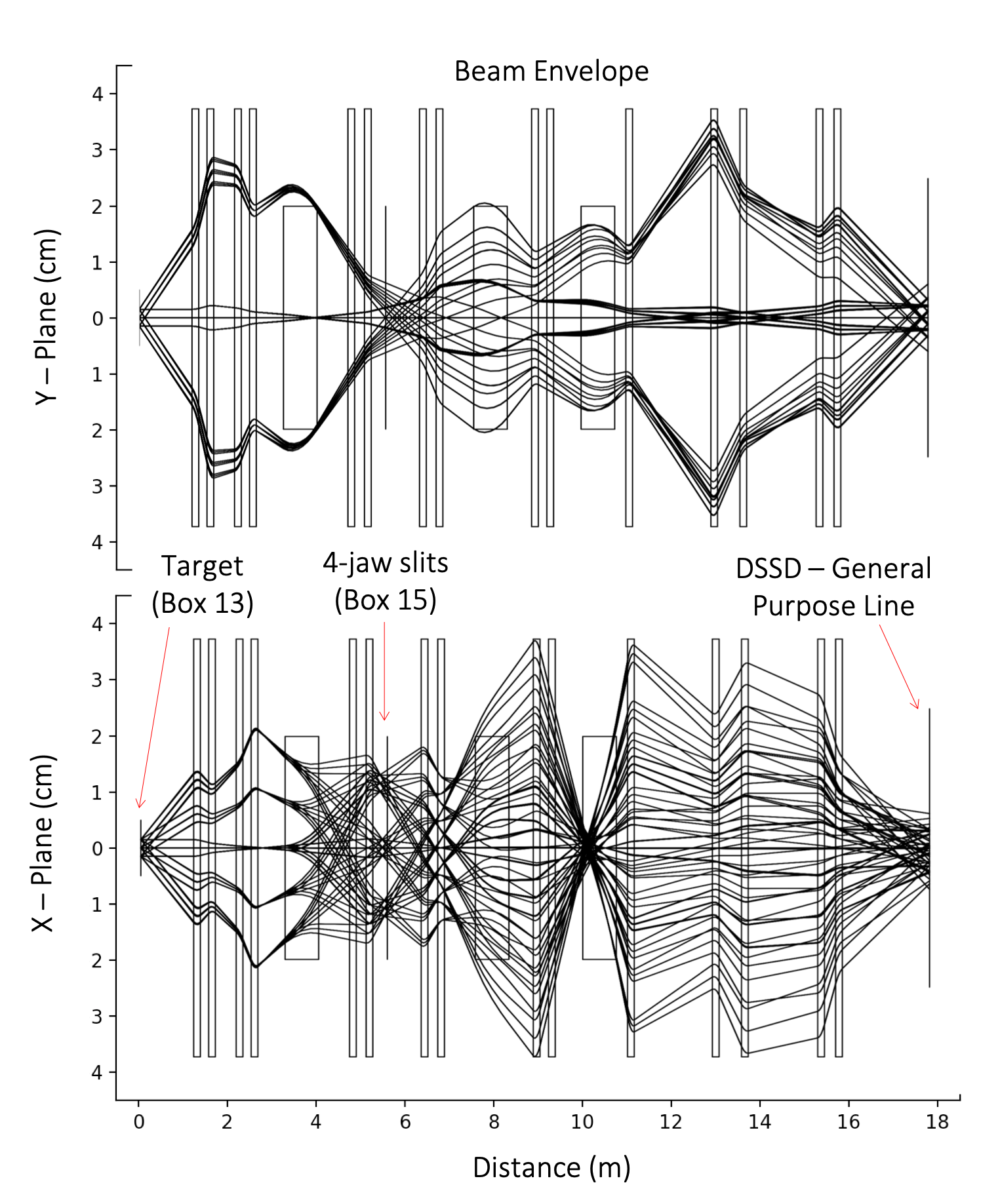}
\caption{Optimized ion optics settings for the ReA3 general purpose line. The beam envelopes were extracted with COSY Infinity (see the text for details). The \textbf{top} panel shows the maximum spread of ion trajectories in the y-plane (vertical), and the \textbf{bottom} panel shows the corresponding trajectories in the x-plane (horizontal). With these settings an energy acceptance of $\pm$1.5\%, and an angular acceptance of $\pm$10~mrad for p$_x$ and p$_y$ is achieved. Furthermore, the x-dispersion versus momentum at the position of the 4-jaw slits system is also maximized. An overview of the beamline can be found in Fig.\ref{fig_setup}.}

\label{fig_cosy}
\end{figure}

\section{Future Applications of the Technique}

\subsection{Ion-Optics Optimization for the ReA3 Beamline}\label{sec_future}

To extract an optimized set of ion optics and facilitate the preparations for future experiments in ReA3, we performed analytical beam-dynamics simulations using COSY Infinity~\cite{Makino:2006}. The calculations were focused on maximizing the energy and angular acceptances of the system downstream of the target (see Fig.~\ref{fig_setup}).

The geometrical characteristics of the general-purpose beamline that we used, set limitations on the maximum acceptance that can be achieved. This acceptance is mainly limited by the diameter of the beam pipes in the section between the two switching magnets in the ReA3 experimental area (see Fig.~\ref{fig_setup}). At this segment of beamline the vacuum pipe diameter is roughly half the size of the rest of the beamline. There exist plans to upgrade this beamline section in the future, whence the limitations we discuss here will be removed. Therefore, for the optimization of the ion-optics that we attempt here, we assume the future beamline arrangement with a uniform diameter of beam pipe throughout.

In Fig.~\ref{fig_cosy}, we show a number of rays defining the beam envelope in the x-plane (horizontal) and y-plane (vertical) using the optimized settings. The graphs show the expected trajectories of the incoming ions, while passing through the various magnetic elements of the beamline. The calculations include terms up to the 3rd order, while the presented ``characteristic rays" (black lines) were extracted based on the maximum angular, lateral, and energy spread that were chosen in the input (see Ref.~\cite{BERG201887}). For the incoming beam, we considered intermediate mass nuclei with A=56 and charge Q=21, while for the beam energy we used E=200~MeV. For the initial beam spot we assumed an area of 3$\times$3~mm$^2$. The starting point of the simulation is the center of the rear flanges (downstream) of diagnostics box 13 where the target is installed. With these conditions, the maximum angular acceptance of the system was found to be 10~mrad in p$_x$ and p$_y$, and the corresponding energy acceptance was $\pm$1.5\%. The pole-tip fields of the quadruple magnets used in the optimized setting are given in Table~\ref{tab:opt_settings}.

Apart from improving the energy and angular acceptance of the system, the optimized settings offer maximum x-dispersion as a function of momentum at the position of the 4-jaw slits system. The calculated momentum dispersion is $\sim$1.5~mm per percent difference in momentum, and has been verified experimentally in ReA3 using various beams, such as $^{85}$Rb and $^{40}$Ar. By maximizing this quantity at the position of the slits, a more precise selection of ions can be achieved based on their momentum. This is particularly useful both for the measurement of reaction cross-sections by improving the separation of unreacted beam from the reaction products, and for optimizing the incoming beams. 
\begin{table}
\centering
\caption{Optimized ion-optics settings for the ReA3 general purpose beamline.  The optimization was done for a reference particle with A=56, Q=21, and E=200~MeV.}
\label{tab:opt_settings}
\begin{tabular*}{0.7\columnwidth}{cc}
 \rule{0pt}{0.5cm} 
 Quadrupole Magnet & Pole-tip Field [kG] \\ \hline
 \rule{0pt}{0.36cm}
QD1272 &   2.55 \\
QD1275	& -2.51  \\
QD1281	& -1.31\\
QD1285	&   2.35\\
QD1307	&  -0.013 \\
QD1310	&  1.80\\
QD1323	&   2.10\\
QD1327	& -1.63 \\
QD1346	& 2.15\\
QD1351	& -0.0050\\
QD1369	& 2.60 \\
QD1388	&  -1.65\\  
QD1395	&  1.01  \\
QD1411	&  1.51   \\
QD1415	&  -1.75 \\
\end{tabular*}
\end{table}

\subsection{(p,n) and (x,n) Experiments with SECAR and Future Spectrometers in the FRIB era.}\label{sec_secar}

\begin{figure}
\includegraphics[width=0.48\textwidth]{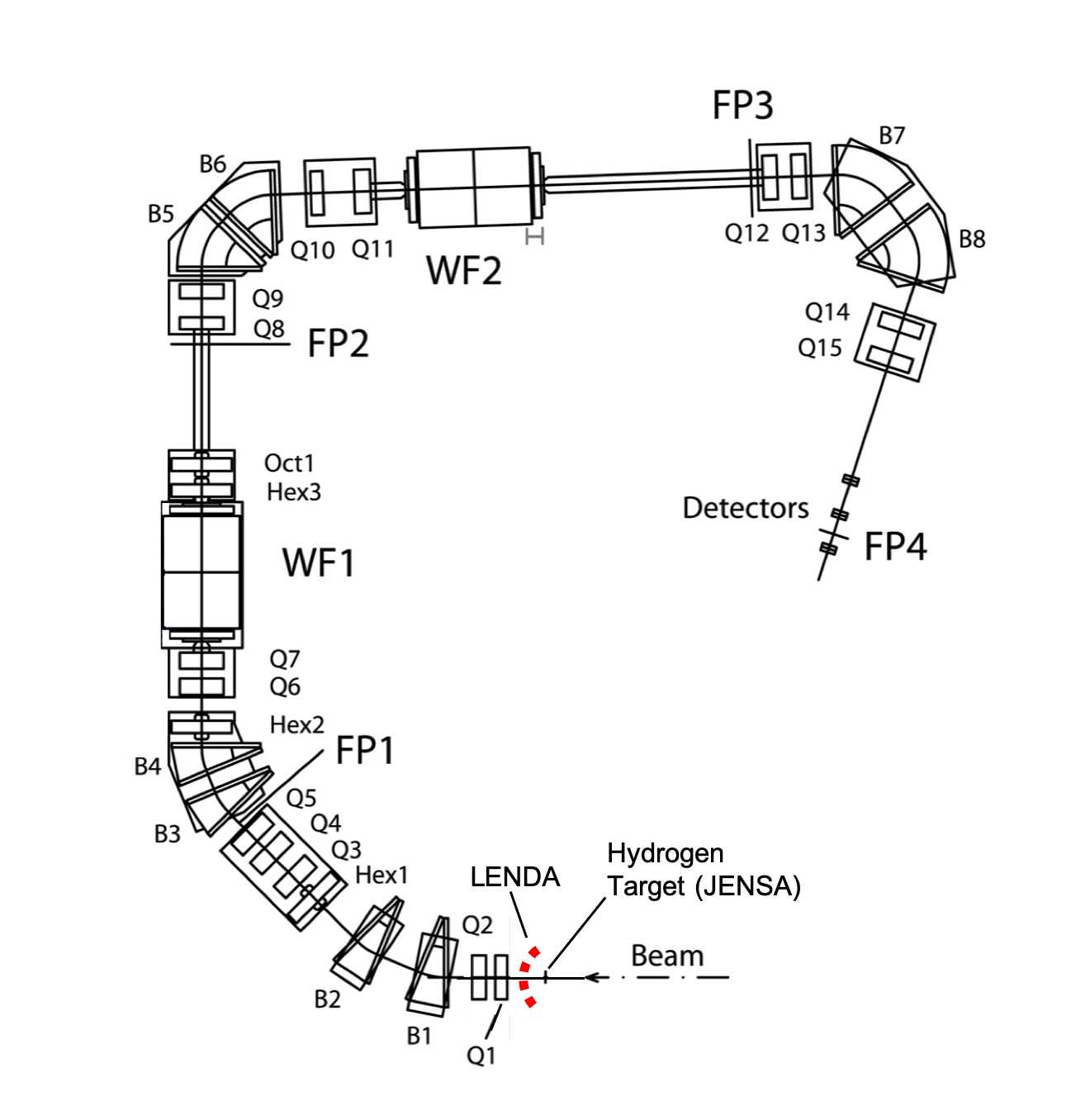}
\caption{ Layout of SECAR recoil separator (modified image from~\cite{Berg:2016}). By integrating the LENDA detector with the rest of system, measurements of (p,n) reactions using the proposed technique can be performed. The heavy reaction products will be detected by a charged-particle detector in coincidence with the neutrons. The charged-particle detector can be placed on any of the various focal planes of the system (FP1, FP2, FP3, FP4)}
\label{fig_secar}
\end{figure}

The proposed technique can also be used in experiments with the SECAR recoil separator. The larger angular and energy acceptance of SECAR (angular acceptances in both transverse directions $\pm$25 mrad, energy acceptance $\pm$3.1\%~\cite{BERG201887}) can increase the transmission of reaction products along the beamline and maximize the efficiency of the measurements. Figure~\ref{fig_secar} shows an overview of the SECAR system combined with the LENDA detector for the measurement of (p,n) reactions. A specially designed frame will allow the arrangement of an optimum number of LENDA bars around the target position. A gaseous hydrogen target will be supplied by the JENSA gas-jet system~\cite{Schmidt:2018gle}, which can provide highly-compressed jet-targets with thicknesses up to $\sim$10$^{19}$ atoms/cm$^2$. In measurements with extremely low counting rates where the areal density of the target atoms is critical, the gas-jet can be replaced by thin plastic foils. The heavy reaction products will be detected either by using a $\Delta$E-E telescope, or an ionization chamber, that can be placed in any of the various focal planes of SECAR (labeled as FP1, FP2, FP3, and FP4 in Fig.~\ref{fig_secar}). By using this configuration, we can take advantage of the unique capabilities of SECAR and perform measurements of (p,n) and other (x,n) type of reactions with radioactive beams using the proposed technique. The detailed analysis and optimization of the SECAR recoil spectrometer for the measurement of (p,n) reaction cross-sections is left for a future work. 

In the last section we provide a rough expected counting rate for such an experiment to make the case for the value of the technique when combined with a recoil separator.
Similar considerations can in principle be entertained for a number of future spectrometers that are being considered for higher-energy experiments with the FRIB reaccelerator, such as, for example, the Isochronous Spectrometer with Large Acceptances (ISLA)~\cite{BAZIN2013319}.  

\subsection{Count Rate Estimates for Future Experiments}

We can make an estimate of the count rates that the technique can achieve in an optimized setup based on the results of the test experiment. We will assume that SECAR is the spectrometer used, but the result can be adapted to any other system as long as we have a known acceptance and transmission. The maximum counting rate for the reaction products during our test was $\sim$15 counts/hour during run-2, and the product transmission was 5.9\%. By assuming an optimized set of ion optics for the ReA3 beamline (see Section~\ref{sec_future}) and a transmission between 80 -- 100\%, the corresponding counting rate is increased to $\sim$230 counts/hour. Taking into account the intensity of the $^{40}$Ar beam and the partial cross-section of the $^{40}$Ar(p,n$_2$) channel ($\approx$15.5~mb), this corresponds to about 7 counts/mb/pnA/s. 
Such a counting rate, taking into account the increase in beam intensity expected from FRIB, means that many unstable isotope studies will be well within the range of experimental capabilities when the FRIB linac comes online.

\section{Summary}\label{sec_summary}

We have developed, tested, and verified an experimental technique for the measurement of (p,n) reactions in inverse kinematics at the ReA3 facility. The technique was verified by measuring the $^{40}$Ar(p,n$_2$)$^{40}$K reaction. The results were compared with the literature data and found to be in excellent agreement within the limits of the experimental uncertainties. The experimental errors of the measured cross-sections varied between 25-30\%. Future improvements of the setup, as well as supplementary measurements of critical quantities such as the charge-state distributions, can reduce the associated errors by an estimated factor of 20\%. Furthermore, by maximizing the transmission of the reaction products through the beamline, count rates of the order of 7 counts/mb/pnA/s can be achieved. The proposed technique can be used for the measurement of (p,n) reactions with heavy radioactive beams. This will allow the study of key (n,p) reactions (reversed reactions) which are relevant to nucleosynthesis via the $\nu$p-process. In the near future, the proposed technique can also be used in experiments with the SECAR recoil separator at the Facility for Rare Isotope Beams.

\section*{Acknowledgements}\label{sec_ack}

We would like to thank the NSCL/FRIB staff members: 
Sam Nash, John Yurkon, Jim Vincent, Zhao Qiang, Tim Embury, Marco Cortesi, Tasha Summers, Daniel Crisp, Laura Pratt, and Alain Lapierre for their invaluable help during the preparation and the execution of the test experiments. Furthermore, we thank the NSCL mechanical design engineers Ben Arend, and Craig Snow, as well as the technicians at the machine shops at NSCL and Central Michigan University, for their contributions on the development of various components for the experimental setup. PG would also like to thank Daniel Alt for his help with the DYNAC software. This material is based upon work supported by the U.S. Department of Energy, Office of Science, under Award Number DE-SC0014285. The authors form CMU would like to acknowledge support by the College of Science and Engineering of Central Michigan University. G.B. acknowledges support by the US National Science Foundation through Grant No. PHY-1068192 and the Joint Institute for Nuclear Astrophysics -- Center for the Evolution of the Elements (JINA-CEE) through Grants No. PHY-0822648 and No. PHY-1430152. The authors from NSCL and MSU would like to acknowledge support by the US National Science Foundation under Cooperative Agreement PHY-156554 (NSCL), PHY-1430152 (JINA-CEE), and PHY-1913554 (Windows on the Universe: Nuclear Astrophysics at the NSCL).

\section*{References}

  \bibliographystyle{elsarticle-num-names} 
  \bibliography{references}

\end{document}